\documentclass{article}

\usepackage{arxiv}

\usepackage[utf8]{inputenc} % allow utf-8 input
\usepackage[T1]{fontenc}    % use 8-bit T1 fonts
\usepackage{hyperref}       % hyperlinks
\usepackage{url}            % simple URL typesetting
\usepackage{booktabs}       % professional-quality tables
\usepackage{amsfonts}       % blackboard math symbols
\usepackage{nicefrac}       % compact symbols for 1/2, etc.
\usepackage{microtype}      % microtypography
\usepackage{graphicx}
\usepackage[numbers]{natbib}
\usepackage{doi}
\usepackage{xcolor}

\begin{document}

\title{Event-based Imaging Velocimetry -- An Assessment of Event-based Cameras for the Measurement of Fluid Flows}

\date{February 18, 2022}	% Here you can change the date presented in the paper title
%\date{} 					% Or removing it

\author{ \href{https://orcid.org/0000-0002-1668-0181}{\includegraphics[scale=0.06]{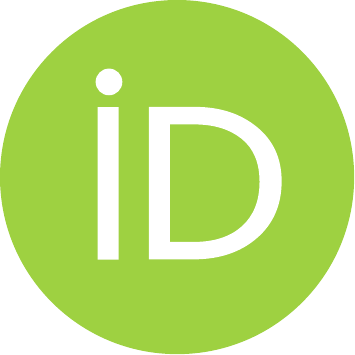}\hspace{1mm}Christian E. Willert} \\
	Engine Measurement Techniques\\
	DLR Institute of Propulsion Technology\\
	German Aerospace Center\\
	51170 K\"{o}ln, Germany \\
	\texttt{chris.willert@dlr.de} \\
	%% examples of more authors
	\And
	\href{https://orcid.org/0000-0003-2709-9664}{\includegraphics[scale=0.06]{orcid.pdf}\hspace{1mm}Joachim Klinner} \\
	Engine Measurement Techniques\\
	DLR Institute of Propulsion Technology\\
	German Aerospace Center\\
	51170 K\"{o}ln, Germany \\
	\texttt{joachim.klinner@dlr.de} 
}

% Uncomment to remove the date
%\date{}

% Uncomment to override  the `A preprint' in the header
\renewcommand{\headeright}{ }
\renewcommand{\undertitle}{under consideration for publication in \emph{Experiments in Fluids}}
\renewcommand{\shorttitle}{EBIV -- Event-based Imaging Velocimetry}

\newcommand{\hlt}[1]{\textcolor{black}{#1}} % to turn it off
\newcommand{\hltred}[1]{\textcolor{red}{#1}}
\newcommand{\todo}[1]{\textcolor{red}{\textbf{ToDo: }#1}}

%%% Add PDF metadata to help others organize their library
%%% Once the PDF is generated, you can check the metadata with
%%% $ pdfinfo template.pdf
\hypersetup{
pdftitle={Event-based Imaging Velocimetry -- An Assessment of Event-based Cameras for the Measurement of Fluid Flows},
pdfsubject={},
pdfauthor={Christian Willert, Joachim Klinner},
pdfkeywords={fluid flow measurement, particle imaging, event-based vision sensing, dynamic vision sensor, neuromorphic imaging, high-speed imaging, time-resolved PIV, particle tracking, PTV},
}

\maketitle

\begin{abstract}
Contrary to conventional frame-based imaging, event-based vision (EBV) or dynamic vision sensing (DVS) asynchronously records binary signals of intensity changes for given pixels with microsecond resolution.
The present work explores the possibilities of harnessing the potentials of event-based vision for fluid flow measurement.
The described implementations of event-based imaging velocimetry (EBIV) rely on the imaging small particles that are illuminated by a laser light sheet which is similar to classical two-dimensional, two-component (2d-2c) PIV with the difference that a continuously operating laser-light sheet is used without modulation of the laser or camera. The moving particles generate continuous time-stamped events on the detector that are later used to infer their velocity using patch-wise processing schemes.
Two flow estimation algorithms are proposed; one uses a ``motion compensation" that maximizes the local contrast, the other is based on a sum-of-correlations approach.
The underlying motion detection schemes along with the complete absence of background signal allows straightforward retrieval of the events associated with individual particles thereby allowing the reconstruction of individual particle tracks.
Alternatively, the event data can be processed with conventional PIV algorithms using images reconstructed from  the event data stream.
The concepts are demonstrated on simple flows in water and air.%
\end{abstract}

\centerline{
\includegraphics[width=0.8\columnwidth]{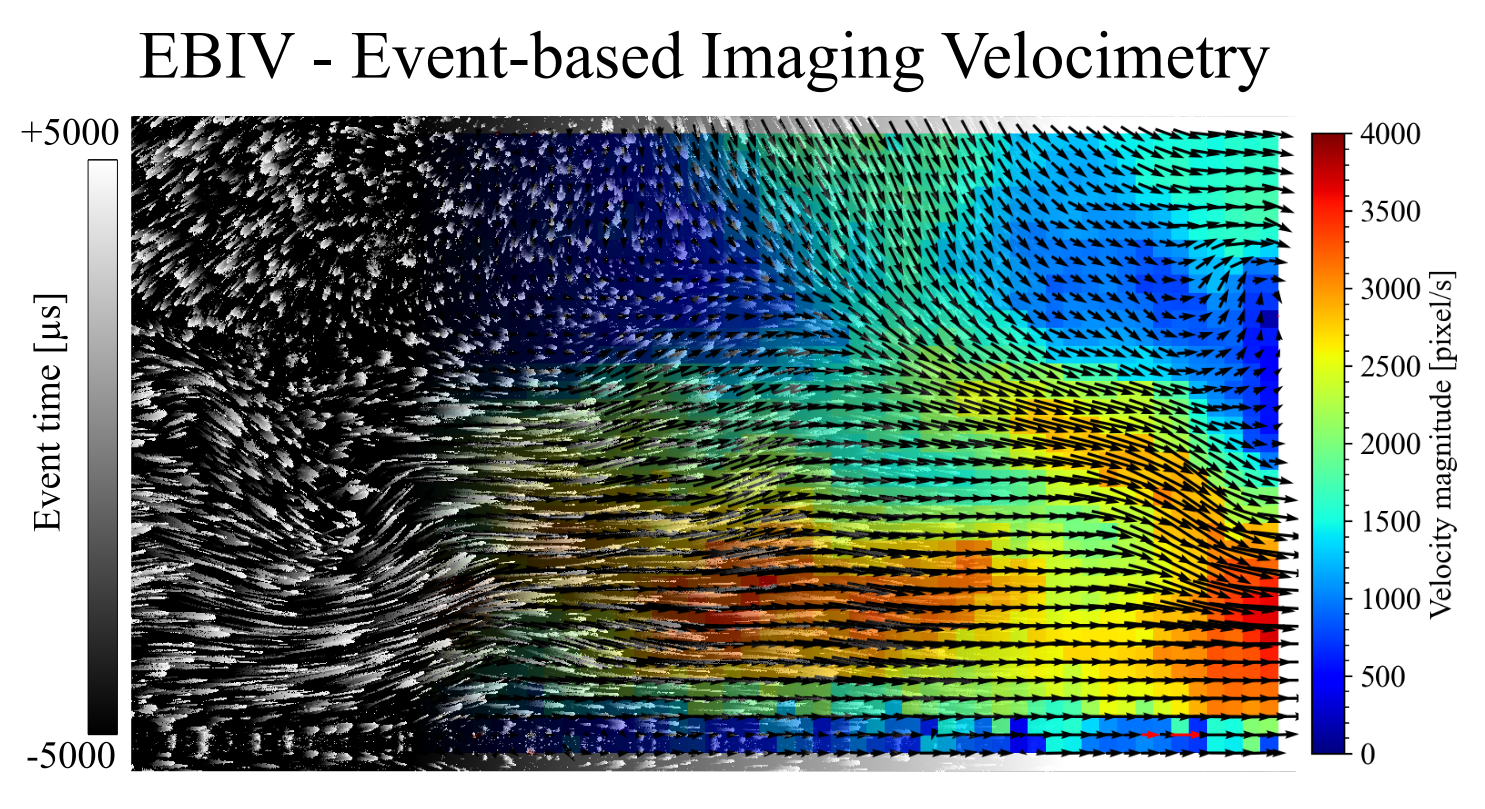}
}

% keywords can be removed
\keywords{fluid flow measurement \and particle imaging \and event-based vision sensing \and dynamic vision sensing \and neuromorphic imaging \and high-speed imaging \and time-resolved PIV \and particle tracking \and PTV}

\section{Introduction}
\label{sec:intro}
\emph{Event-based vision} (EBV), also termed \emph{dynamic vision sensing} (DVS) or neuromorphic imaging, is a new upcoming field within the field of computer vision.
Contrary to conventional frame-based imaging, EBV only records changes of image intensity (i.e. contrast changes) on the pixel level, triggering a positive event ($+1$) for increasing intensity and a negative event ($-1$) for a decreasing intensity change. The typical threshold of the intensity-change trigger is on the order of 20\% but can be fine tuned.
As the pixels on the detector respond individually, the events appear asynchronously throughout the detector area resulting in a continuous stream of data, with each event datum $E_i = E_i(\mathbf{x},t,p)$ consisting of pixel coordinates $\mathbf{x_i} = (x_i,y_i)$, a time stamp $t_i$ and a polarity $p_i \in \{+1,-1\}$ indicating the sign of the intensity change

Original prototype development of the technology dates back to work Mahowald and colleagues at the California Institute of Technology in the 1990's and was initially referred to as \emph{silicon retina} \cite{Mahowald:1992} as the intention of the imaging approach was to mimic the function of the eye's retina.
First practical implementations of EBV resulted from work at University of Zurich as well as ETH Zurich around 2008 \cite{Lichtsteiner:2008,Posch:2014}.
In recent years, several ready-to-use cameras and sensor evaluation kits based on the EBV technology have become commercially available.
This has broadened the range of applications as testified in a steadily increasing number of publications (see e.g. \cite{GIT:EBVrefs}) and also has made the present feasibility study possible.
For a recent review of the technology and underlying concepts the reader is referred to \cite{EBVreview:2022}.

The application of EBV for the visualization and measurement of fluid flows is by no means new.
Initial work was performed by Drazen et al. \cite{Drazen:2011} on particle tracking velocimetry (PTV) of dense particles in a solid–liquid two-phase pipe flow using an EBV sensor of 256$\times$256 pixels and CW laser (5W) illumination.
Ni et al. \cite{Ni:2012} used an EBV array of 128$\times$128 elements to demonstrate micro-particle tracking ($\mu$PTV) with 12 $\mu$m microspheres and were able to detect Brownian motion.
First PTV measurements in an air flow were performed by Borer et al. \cite{Borer:2017} using three synchronized EBV cameras (128$\times$128 pixels) to track helium-filled soap bubbles (HFSB) in volumes up to about 1\,m side length using white light LED arrays for illumination. The flow was only sparsely seeded allowing individual particles to be tracked throughout the volume with final data sets containing up to O(1\,000 - 10\,000) tracks.
More recently, Wang et al. \cite{Wang:2020} implemented a stereoscopic EBV system using two event cameras with 480$\times$360 elements. They reconstructed three-dimensional tracks using 2-d tracking results from each of the cameras. Their flow experiment consisted of a small hexagonal cell with stirrer inducing a swirling flow containing O(100 $\mu$m) polystyrene spheres.

The present work aims at assessing the potential of EBV for more ``traditional" planar 2d-2c flow measurement, ideally being able to extend the findings to full 3d-3c measurements. In particular, much higher seeding densities than in previous studies are investigated and are made possible through recent advances in EBV hardware development.
Two commercially available event-based imaging cameras are used for the investigations with their specifications provided in Table~\ref{tbl:ebi_cameras}.
While the data presented in the remainder of the article is acquired with the high resolution camera both cameras produce event data streams of similar quality with a direct comparison beyond the scope of the material presented herein.
Measurements are demonstrated in both water and air using the same seeding particles and densities as used for conventional PIV. To the best of the authors' knowledge, EBV based fluid flow measurements so far have not been demonstrated at considerably higher particle image densities.

The paper is structured as follows: first, recently introduced motion detection schemes for EBV are briefly described with regards to their suitability for fluid velocity estimation, from which two approaches are selected to be adapted to the present application.
Sec.~\ref{sec:waterflow-exp} introduces measurements on a simple water flow and provides illustrative, real-world examples of both the acquired event image data and the application of the motion detection schemes.
In Sec.~\ref{sec:performance} the performance of the algorithms is assessed using a known flow (turntable) and scatter plots of a turbulent flow to highlight possible systematic sources of error.
The current limits of EBIV are then tested using an air flow with a wide range of velocities and scales (Sec.~\ref{sec:airflow}).
The discussion section provides an overview of both the advantages and disadvantages of the EBIV technique, followed by concluding remarks concerning possible improvements and further applications.

\begin{table}
  \caption{Specifications of cameras used for the present study}\label{tbl:ebi_cameras}
  \centering
\begin{tabular}{|l|c|c|}
  \hline
  Camera Model & \textbf{SilkyEvCam EvC3A} & \textbf{Evaluation Kit 2 - HD} \\[2pt]
  \hline
  Provider & Century Arks & Prophesee \\
  Sensor & Gen3, PPS3MVCD & Gen4.1, HD CIS-BSI \\
    & (Prophesee) & (back-side illuminated, \cite{Finateu:2020}) \\
  Array size (W$\times$H) & 640$\times$480 & 1280$\times$720 \\
  Pixel size & 15$\times$15 & 4.86$\times$4.86 \\
  Fill factor & 25\% & $>$77\% \\
  Nominal contrast threshold &  25\% & 25\% \\
  Minimum detectable contrast change$^1$ &  12\% & 11\% \\
  (50\% response)  & & \\
  Dynamic range & $>$120\,dB & $>$110\,dB \\
  Latency (spec-sheet) & 200\,$\mu$s & 220\,$\mu$s \\
  Latency$^1$ & 40 - 200\,$\mu$s & 20 - 150\,$\mu$s \\
  Time-stamp resolution & 1\,$\mu$s & 1\,$\mu$s \\
  Bandwidth (events/s) & $\approx$550 M & 1066 M \\[2pt]
\hline
\end{tabular}\\
\small{$^1$) Values sourced from Table~1 in \cite{EBVreview:2022}}
\end{table}

\section{Event-based vision and optical flow}
\label{sec:opticalflow}
Within the computer vision community considerable research has been performed in recovering the optical flow from conventional frame-based image sequences and, more recently, event-based image data.
In part, these efforts are aimed at transferring the concepts from frame-based imaging to event-based imaging, but have only partially been successful, which is mostly due to the fact that the well established optical flow approaches rely on  brightness constancy \cite{HornSchunk:1981,LucasKanade:1981}. In event-based imaging, image brightness per se is not available, such that some of the proposed optical flow algorithms rely on reconstructed image intensity as an intermediate quantity \cite{Benosman:2012} or recover the brightness field along with the optical flow \cite{Bardow:2016}.

The output of the optical flow algorithm can be classified as ``dense" or ``sparse", respectively providing motion estimates for every pixel or on selected pixels only. Event image data produced by discrete, small particles initially only provide sparse motion fields where the sparseness directly corresponds to the source density, namely, the imaged particles. To obtain a dense field (for every pixel) requires a smoothness regularization that results in a continuous variation of the velocity estimate within the sampled volume.
Properly converged, the difference between the velocity estimate and the motion of the individual particles approaches zero.

\subsection{Suitable processing algorithms for EBIV}
\begin{figure}[htb]
    \centering
    \includegraphics[width=0.9\columnwidth]{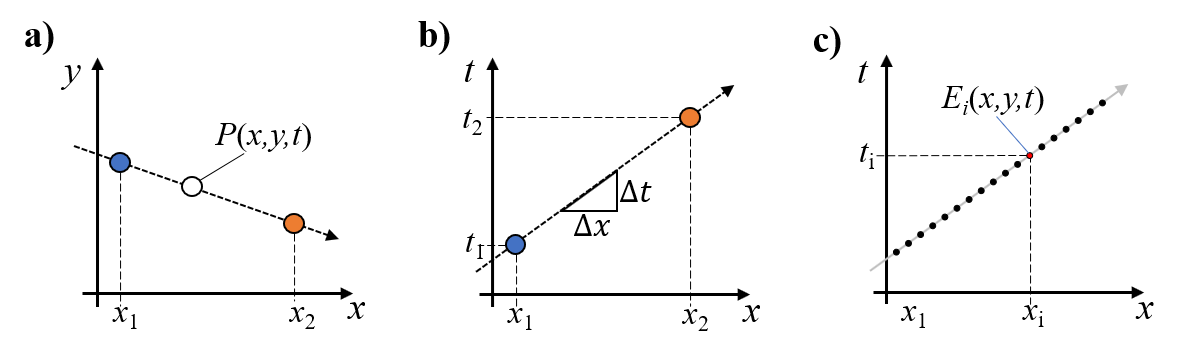}
    \caption{Illustration a single particle $P$ moving at a fixed velocity $V = (v_x,v_y)$ in 2d space (a), in $x-t$ space (b) and trail of events produced by the particle in $x-t$ space (c).}
    \label{fig:particle_motion}
\end{figure}

\begin{figure}[htb]
    \centerline{
    \includegraphics[width=0.36\columnwidth]{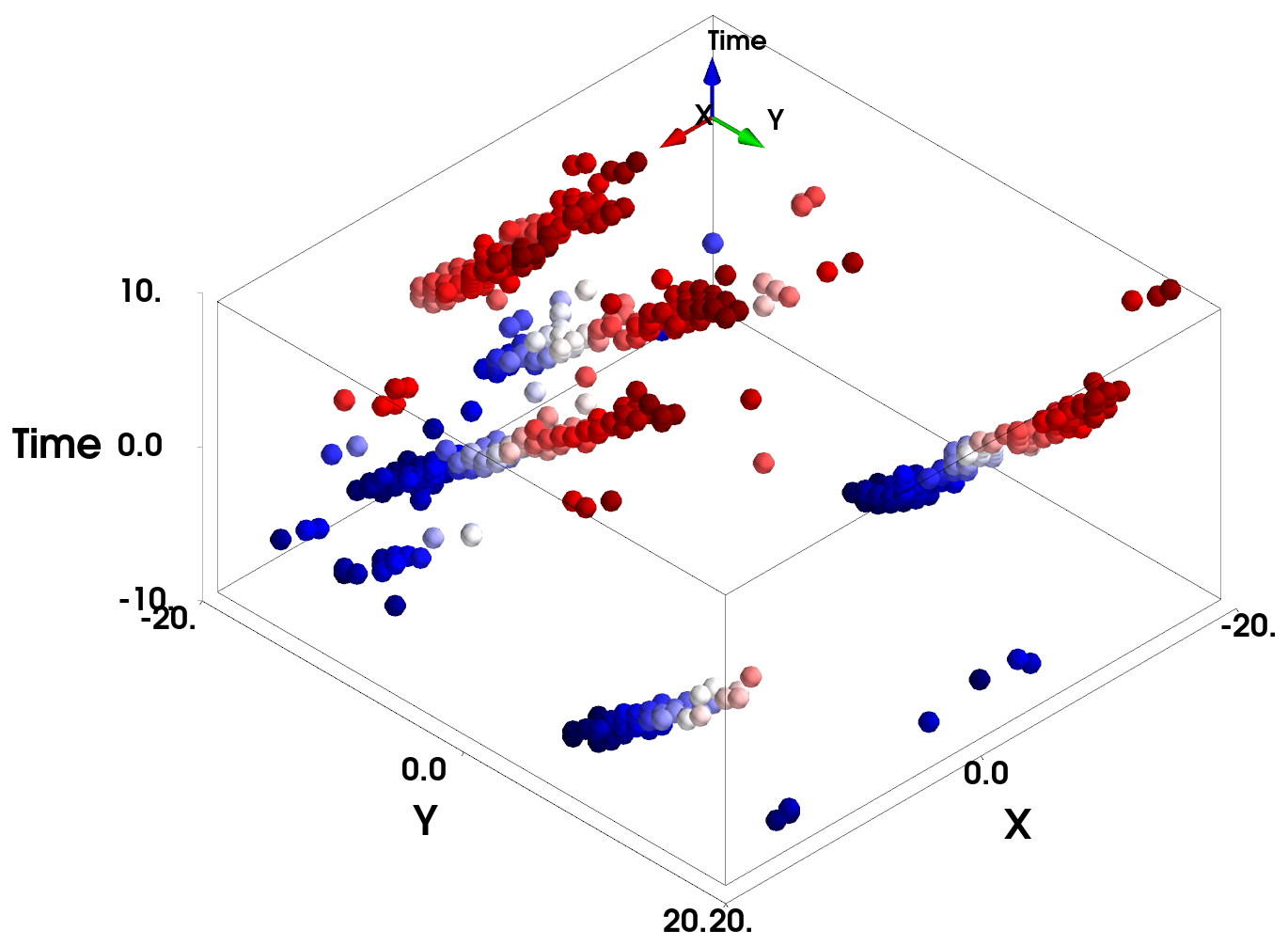}
    \includegraphics[width=0.28\columnwidth]{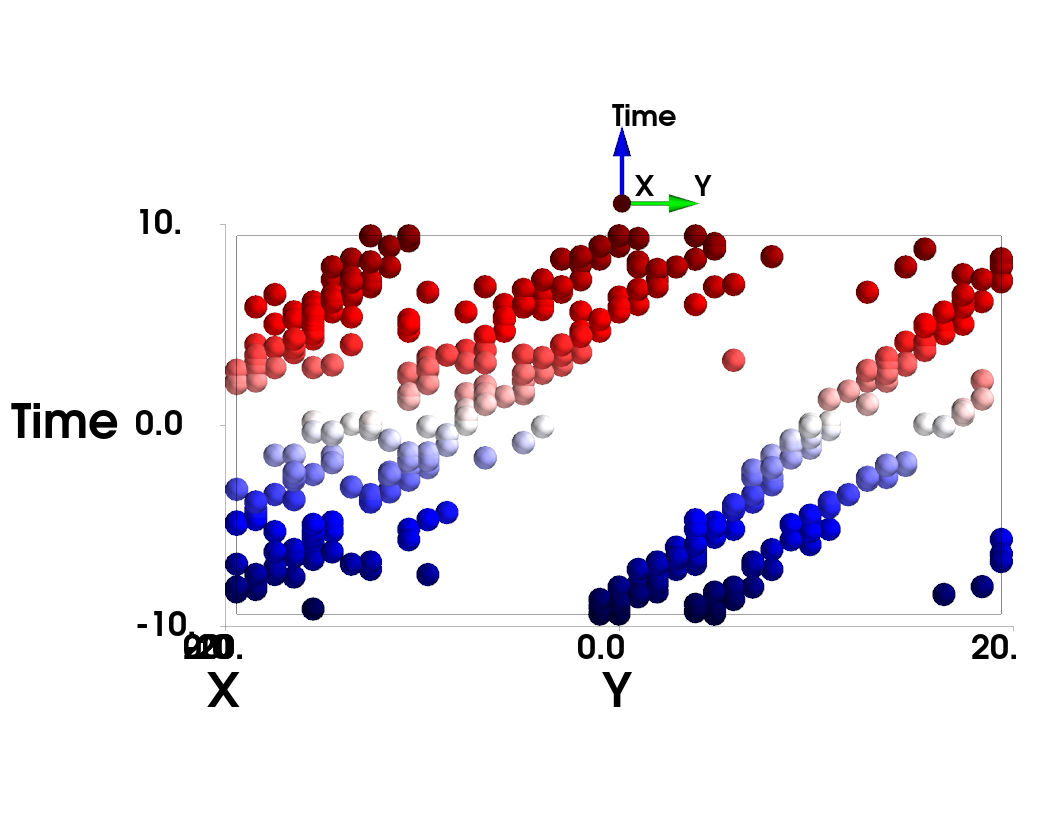}
    \includegraphics[width=0.35\columnwidth]{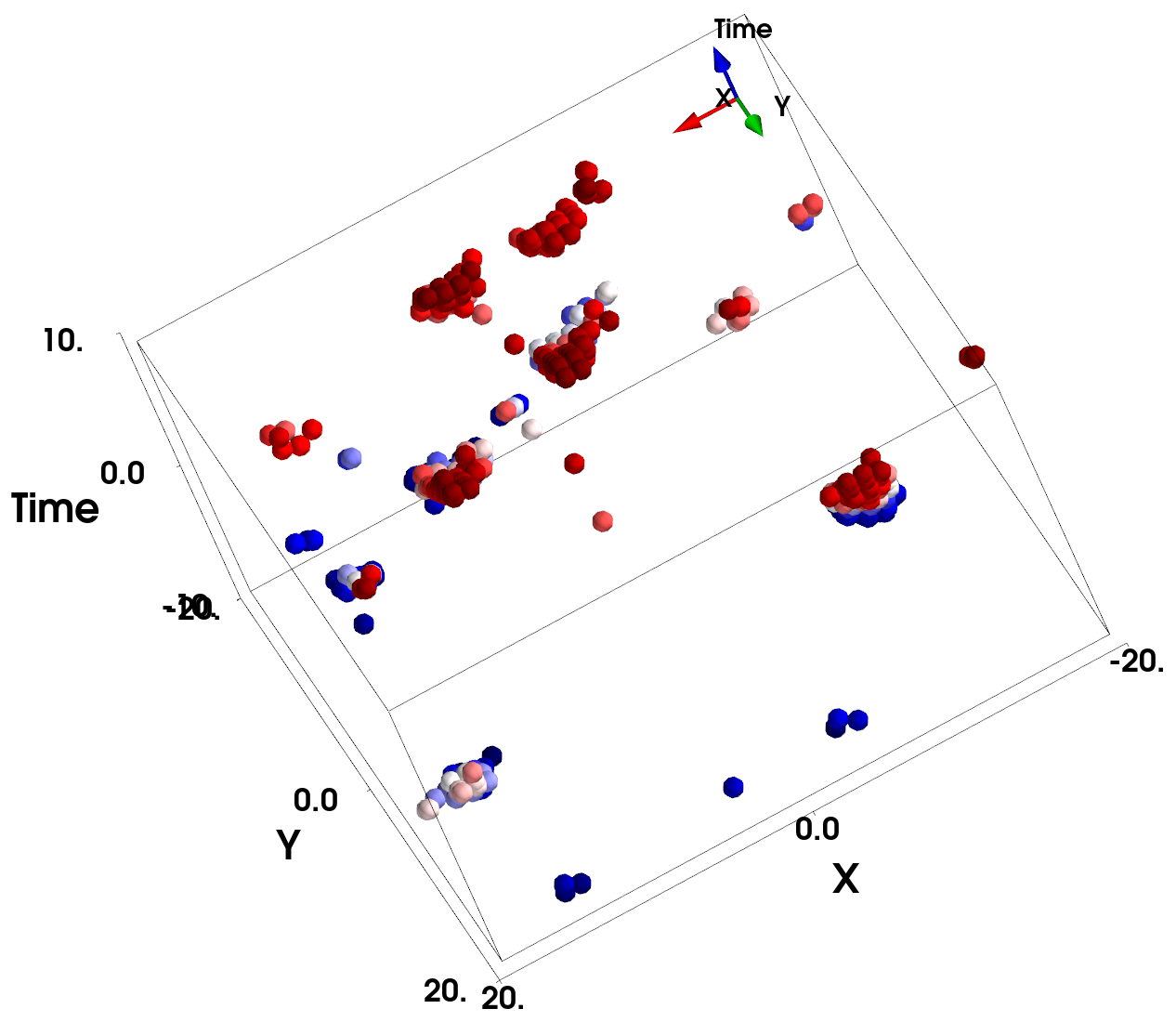}}
    \caption{3-d rendition of a sub-volume of events of size 40$\times$40~pixels and 20\,ms duration. Blue color coding marks past events while red indicates later events. Left: isometric view; middle: view aligned with x-axis; right: view direction chosen to align events into clusters.}
    \label{fig:3d-view}
\end{figure}

To illustrate the concepts of motion estimation, Figure~\ref{fig:particle_motion} shows the motion of single particle $P$ both through planar space ($\mathbf{x} = (x,y)$) as well as in the space-time domain.
Assuming a locally constant velocity, a first order estimate of the particle velocity is given by the slope of the particle path in the space-time domain, which is the sought-after quantity for which a suitable estimation algorithm has to be found.
Figure~\ref{fig:3d-view} provides a 3d rendition of actually recorded events of particles images in a water flow (see Sec.~\ref{sec:waterflow-exp}).

The approach chosen here follows the concepts used in the computer vision community for optical flow estimation.
Ideally, the optical flow estimation algorithm should not require to explicitly reconstruct image intensity, that is, it operates only by using the spatio-temporal resolved events $E_i = E_i(\mathbf{x},t,p)$.
Event data generated from imaged objects (i.e. cars, people, trees, etc.) typically consist of broad ``swaths" of events that, when aligned, result in sharp edges.
By fitting a plane in $x-y-t$ space to these events, Benosman et al.~\cite{Benosman:2014} show that optical flow data can be retrieved but the approach has its limitations on more complex scenes \cite{Rueckauer:2016}.
Imaged particles, on the other hand, typically only cover a few pixels when properly focussed.
In fact, for the PIV technique an optimal particle image has a size of 2-3 pixels ($e^{-2}$ diameter \cite{PIVbookAdrianWesterweel,PIVBookRaffel:2018}). This makes a plane fitting approach challenging.
Another approach relies on the convolution of the event data with separable space-time filter kernels to estimate the motion from the maximum response \cite{Brosch:2015}.

\begin{figure}[htb]
    \centering
    \includegraphics[width=0.9\columnwidth]{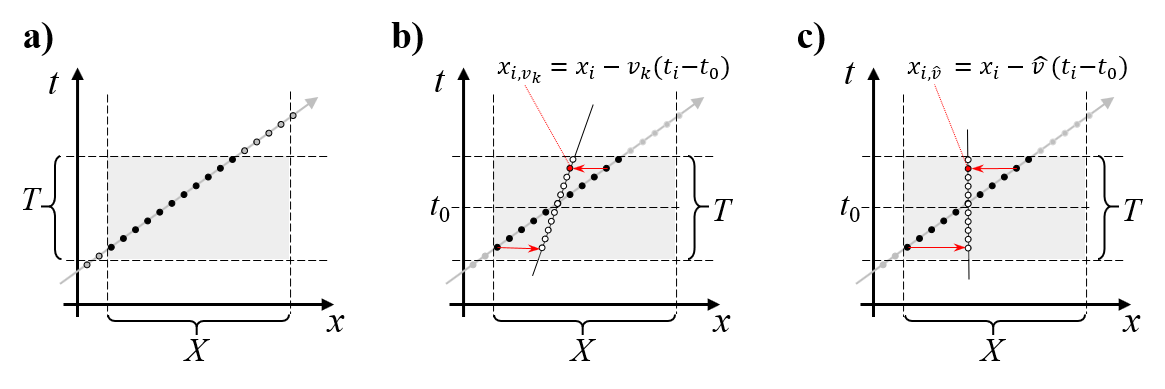}
    \caption{Principle of patch-wise motion compensation; a) events are sampled in volume of spatial dimension $X$ and time interval of $T$, b) motion compensation is  performed iteratively using velocity candidates $v_k$; c) optimal velocity $\hat{v}$ is found when events line up at a constant position $\mathbf{x}_{i,\hat{\mathbf{v}}}$. }
    \label{fig:motion_comp1}
\end{figure}

At present, one of the most suitable approaches for EBIV seems to be the so called ``motion compensation" introduced by Gallego et al.~\cite{Gallego:2018}.
The idea behind motion compensation is to align the events such that they line up using a linear skewing (warping) of the sampled $x-y-t$ sub-volume.
Conceptually this can be visualized by choosing a viewing direction onto a $x-y-t$ sub-volume that makes the events in this volume appear to be lined up, as illustrated in Figure~\ref{fig:3d-view}.
Event tracks produced by individual particles appear to converge into a small cluster. The viewing direction corresponds to a vector $V = [dx/dt,dy/dt]$ that ``warps" the events such that they converge (cf. Figure~\ref{fig:motion_comp1}c).
The motion compensation is applied for sub-volumes of size $X, Y$ in space and $T$ in time (block matching \cite{Liu:2018}) assuming constant (linear) motion within the sampled volume.
Optimal alignment of the particle image events is achieved when the variance in the warped domain is maximized, that is, the image contrast is maximized once the motion blurring has been removed.

An alternative method of retrieving the optical flow from event data is proposed by Nagata et al.~\cite{Nagata:2021} who claim that their method is more robust in comparison to the previously described iterative variance-based motion compensation approach.
In their case the local optical flow is obtained from two sub-volume at the same location slightly offset in time $\tau$ and measures the consistency of the time-surfaces formed by the events in $x-y-t$ space between the two samples.
The minimization-based algorithm relies on spatial derivatives of the time surface and requires a smoothness regularization term to stabilize the minimization.

\begin{figure}[htb]
    \centering
    \includegraphics[width=0.9\columnwidth]{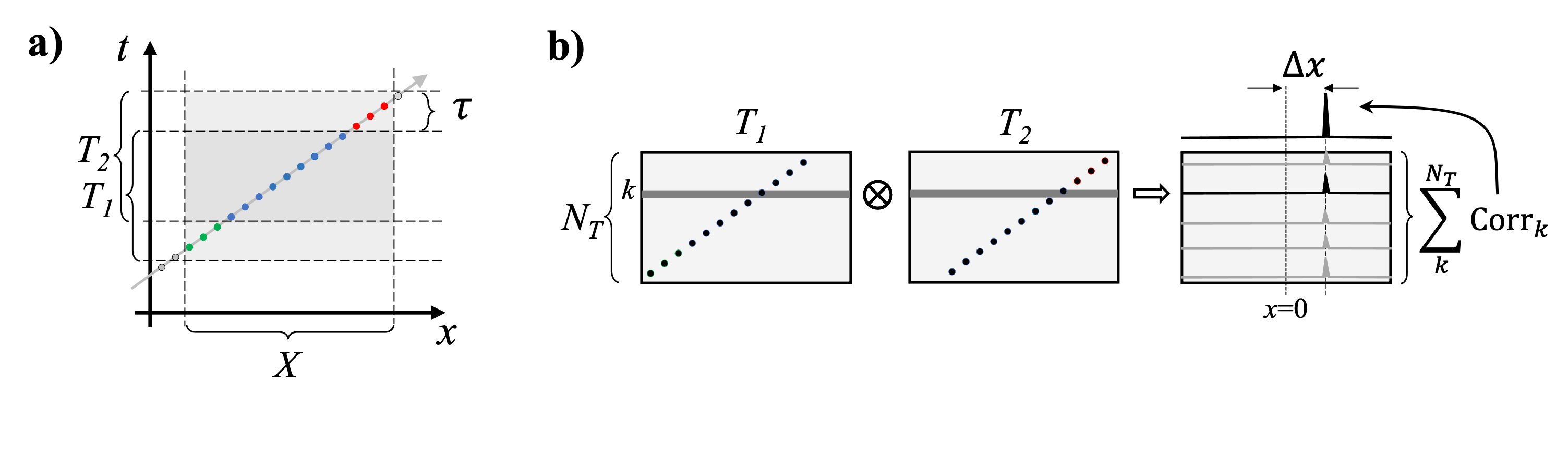}
    \caption{Principle of patch-wise velocity estimation using a temporal offset; a) events are sampled in two volumes of equal spatial dimension $X$ for a time interval $T$, but separated by a time delay of $\tau$, b) mean displacement $\Delta x$ is determined by summing $N_T$ separate cross-correlations of events for different time slots $t_k$. Mean velocity is given by ratio $\Delta\mathbf{x}/\tau$.}
    \label{fig:crosscorr_method}
\end{figure}

For the present study the algorithm by Nagata et al.~\cite{Nagata:2021} has been not implemented.
Instead, we propose to use their concept of sampling the $x-y-t$ volume with two sub-volumes separated by time $\tau < T$ and apply a motion estimation scheme that is more suited for particle image data.
The basic principle is outlined in Figure~\ref{fig:crosscorr_method}: the two sub-volumes are subdivided into $N_T$ equal time-slices of duration $T/N_T$.
Events within each time-slice are then combined and result in $N_T$ planes of events within $x-y$ space for each sample.
Next, the cross-correlation is computed for each pair of event planes.
Finally, the $N_T$ cross-correlation planes summed to form a combined a single combined correlation plane.
The position of the maximum correlation value with respect to the origin represents the most likely particle displacement within the sampled volume.
Division of the correlation peak displacement by the sample time-offset $\tau$ yields a first-order estimate of the ``pixel velocity" within the sampled sub-volume.

For both approaches the motion compensation is performed symmetrically around time $t_0$ making them central difference schemes which are inherently second-order accurate \cite{WereleyMeinhart:2001}.
To improve sub-pixel performance, 3-point parabolic fits or 3-point Gauss fits are respectively applied in X and Y direction using the neighboring values of the maximum variance or maximum correlation.

In order to retrieve velocity data across the field of view, the simplest algorithm involves a patch-wise processing very similar to conventional PIV. The sample size is fixed throughout the field of view and the interrogation is performed on a regular grid regardless of the local particle image (e.g. events) concentration.
For the motion compensation approach a sufficient range of ``pixel velocities" needs to be defined a priori to ensure the capture of the variance peak. Similarly, an adequate time-offset $\tau$ has to be chosen for the sum-of-correlation scheme to capture events of the same particle(s) in both $x-y-t$ sub-volumes.

The logical extension of the single-pass approaches are \emph{grid-refining schemes} that start coarse and iteratively move into finer grid resolutions, both spatially and with increasingly finer velocity resolution.
Here deformation schemes that allow a continuous variation of the velocity estimate within the sampled volume can be directly imported from state-of-the-art PIV processing algorithms.

When converged, the iterative, patch-wise particle motion detection scheme effectively results in clusters of pixels that are associated with the motion of single particle (see e.g. 3rd column in Figure~\ref{fig:real_variance_processing}). Therefore, it is a (nearly) trivial step to single out the events for an individual particle image from the original event data set. These events can then be used to recover the track of the given particle in both space and time using an adequate fitting scheme (see e.g. methods used for 3-d Shake-the-box Lagrangian tracking \cite{Gesemann:2016,Schanz:2016}).

In the following section implementations of both the motion compensation algorithm and the sum-of-correlation  scheme are used to recover the flow field of a simple water flow experiment.

\section{EBIV Applied to a Water Flow Experiment}
\label{sec:waterflow-exp}
\begin{figure}[tb]
\centerline{%
    \includegraphics[width=0.46\columnwidth]{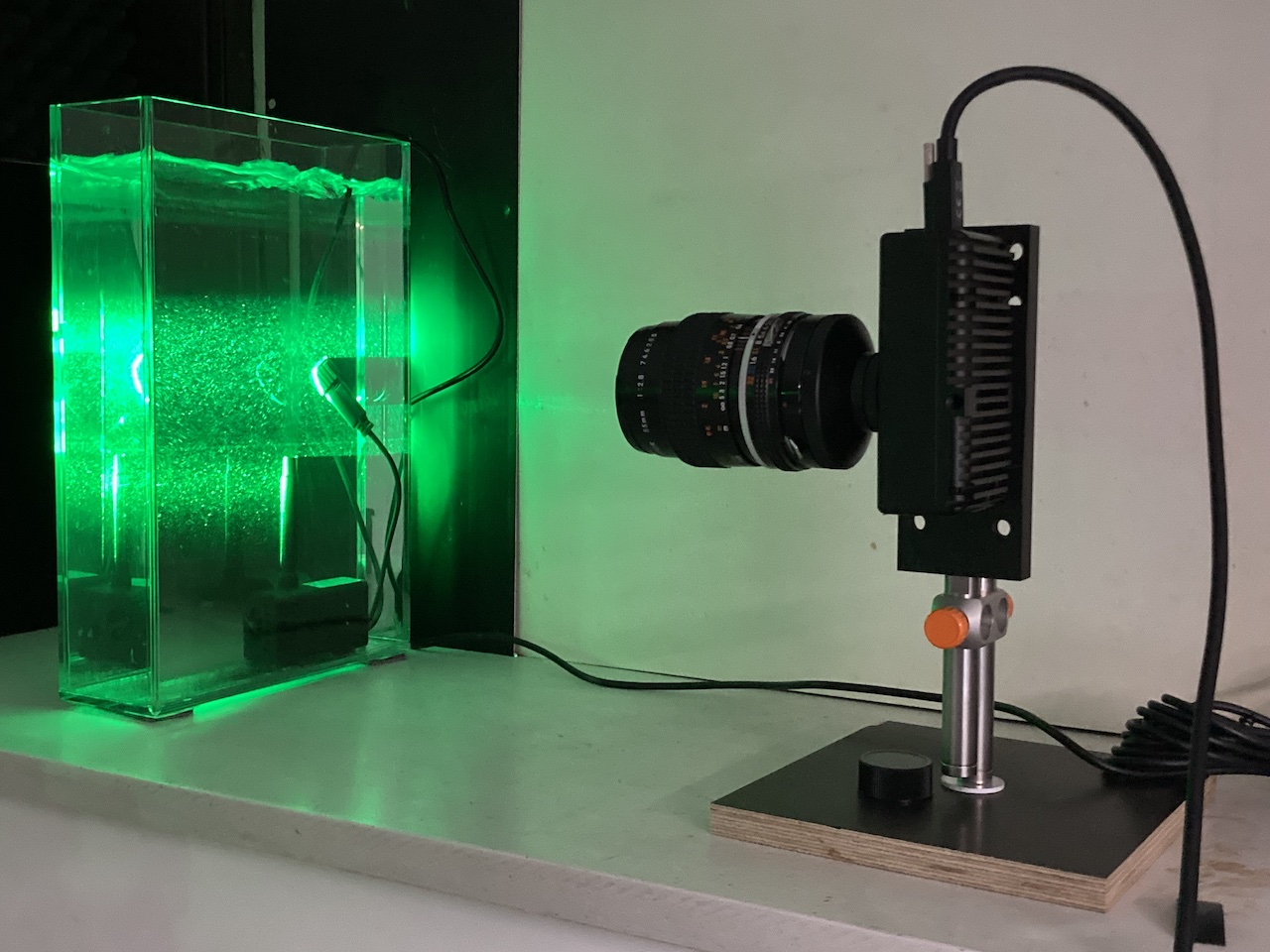}\hfill
    \includegraphics[trim=135 0 0 0, clip=true, width=0.52\columnwidth]{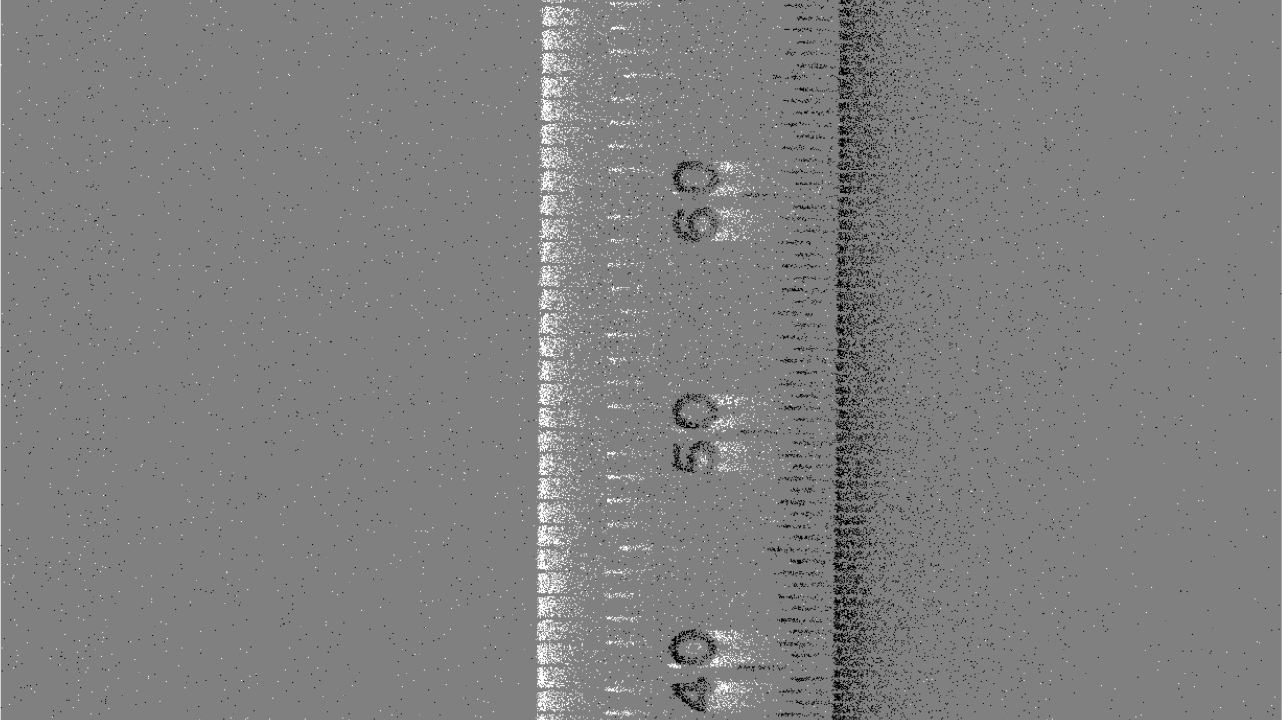}%
    }
    \caption{Left: Photograph of a simple water flow EBV imaging setup in the ``home-lab" of the first author involving a small water tank, laser light sheet and event-based imaging camera.
    Right: Image of a ruler moving sideways recorded by the event camera illustrating the event generation within a period of 10\,ms (white pixels are positive (``on") events, black pixels are negative (``off") events).}
    \label{fig:photo_waterflow}
\end{figure}

A water basin of about $400\times 250 \times 60\,\textrm{mm}^3$, is seeded with nearly neutrally buoyant  silver-coated, hollow glass spheres of about $10\,\mu$m diameter.
A small pump on the side introduces a globally recirculating flow with underlying turbulent structures (see Figure~\ref{fig:photo_waterflow}).
The particles are illuminated by a $\approx$100\,mm wide laser light sheet with about 1\,mm waist thickness created with  a laser (Kvant Laser, 4W max., $\lambda$ = 520\,nm) operating in continuous mode (CW).
The light sheet is induced through the vertical side of the tank and the camera is rotated by 90$^\circ$ such that the glass wall appears at the bottom of the recorded images.

\begin{figure}[tb]
\centering
    \includegraphics[width=\columnwidth]{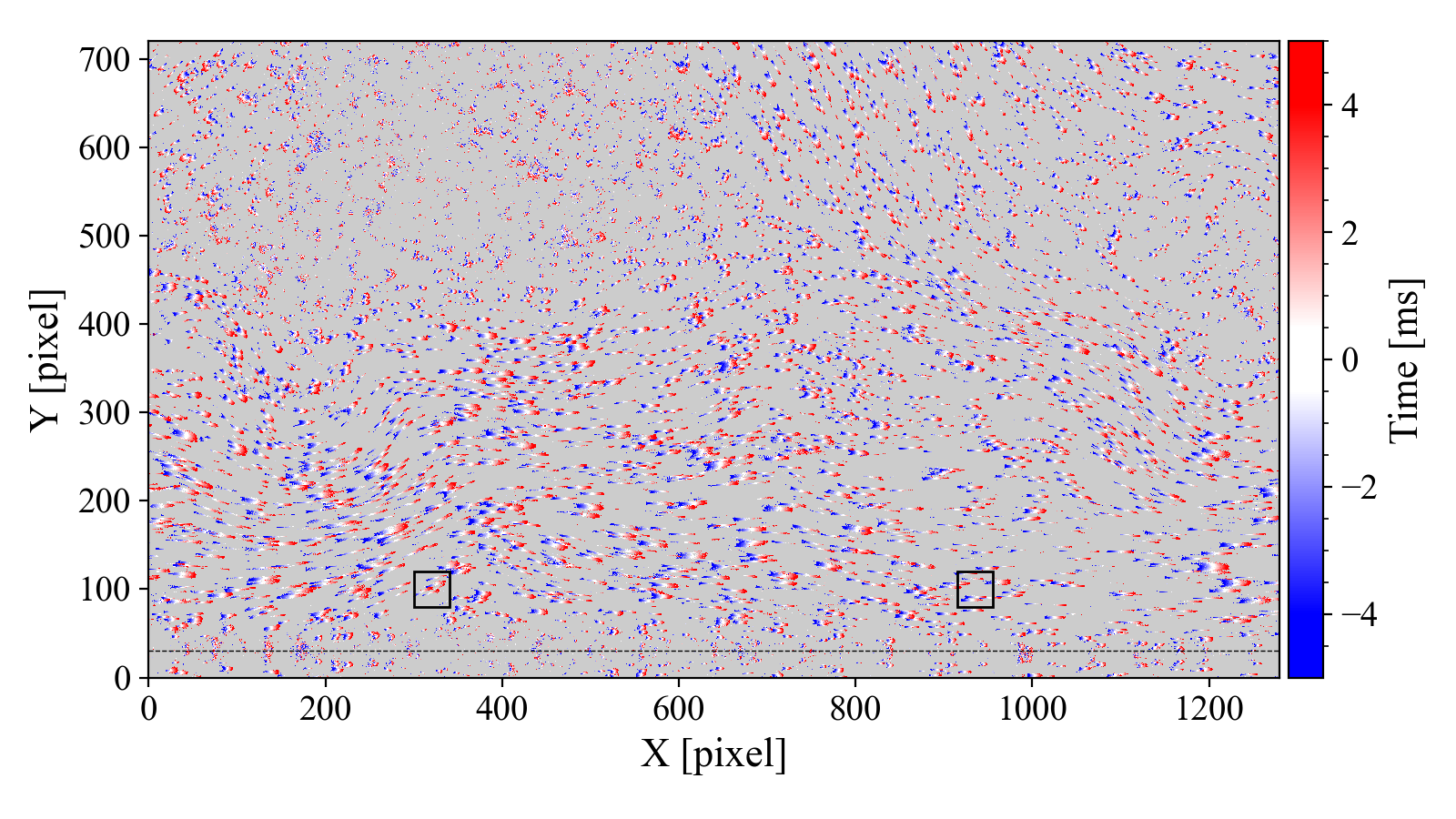}
    \caption{Image constructed from 10\,ms worth of positive events triggered by particles moving in a tank of water and illuminated by a CW laser light sheet. The mean flow is from left to right.
	The black rectangles indicate sample areas of 40$\times$40\,pixels used in the following figures.
	The thin line near the bottom indicates the position of the glass tank wall which is essentially invisible due to the absence of intensity events.}
    \label{fig:events_image}
\end{figure}

Figure~\ref{fig:events_image} shows an ``image" of events compiled from a time-interval (10\,ms) of recorded events sampled on a field of view (FOV) of 54$\times$30 mm$^2$ at a magnification of 42.3\,$\mu$m/pixel. The data was acquired at a data rate of 2$\times$10$^7$ events per second which is two orders of magnitude below the maximum supported by the camera.
The reconstructed image of Figure~\ref{fig:events_image} contains a total of about 130\,000 events that are color-coded using their respective time of occurrence. Events marked blue occurred are in the ``past" while red events occur in the ``future" with respect to the reference time $t_0 = 0$, for which events are coded in white.
Note, that the wall through which the light sheet is introduced is not at all visible, which is a direct result of the lack of temporal intensity-variations at the glass-water interface. A dashed line in Figure~\ref{fig:events_image} indicates the approximate position of the wall.
Animations of the raw event image data and recovered velocity fields are provided in the supplementary material.

\begin{figure}[htb]
\centering
    \includegraphics[width=0.495\columnwidth]{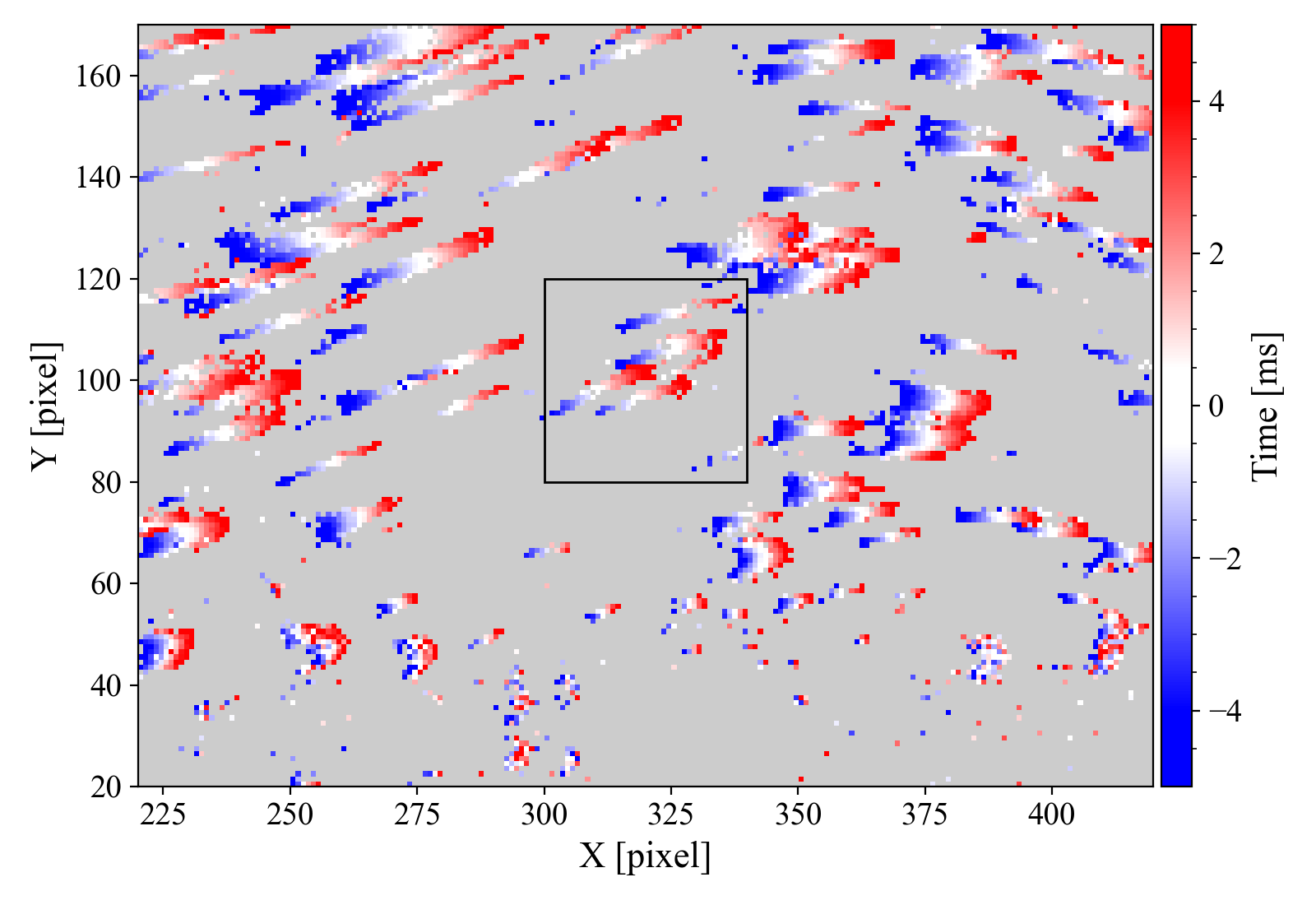}
    \includegraphics[width=0.495\columnwidth]{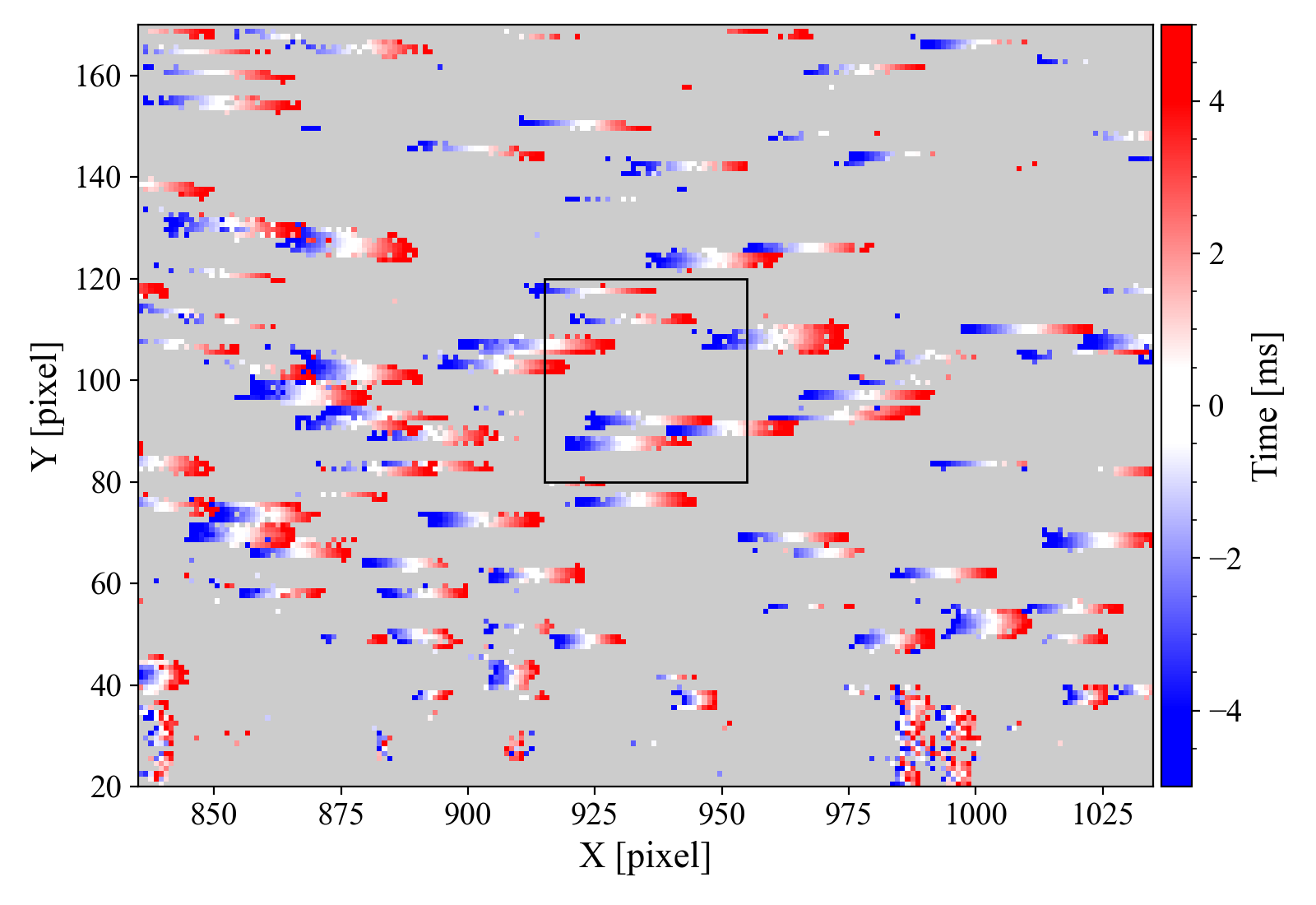}
    \caption{Zoomed-in portions of Fig.~\protect\ref{fig:events_image}.
	The black rectangles indicate sample areas of 40$\times$40\,pixels.}
    \label{fig:events_image_detail}
\end{figure}

\begin{figure}[htb]
    \centering
    \includegraphics[width=0.70\columnwidth]{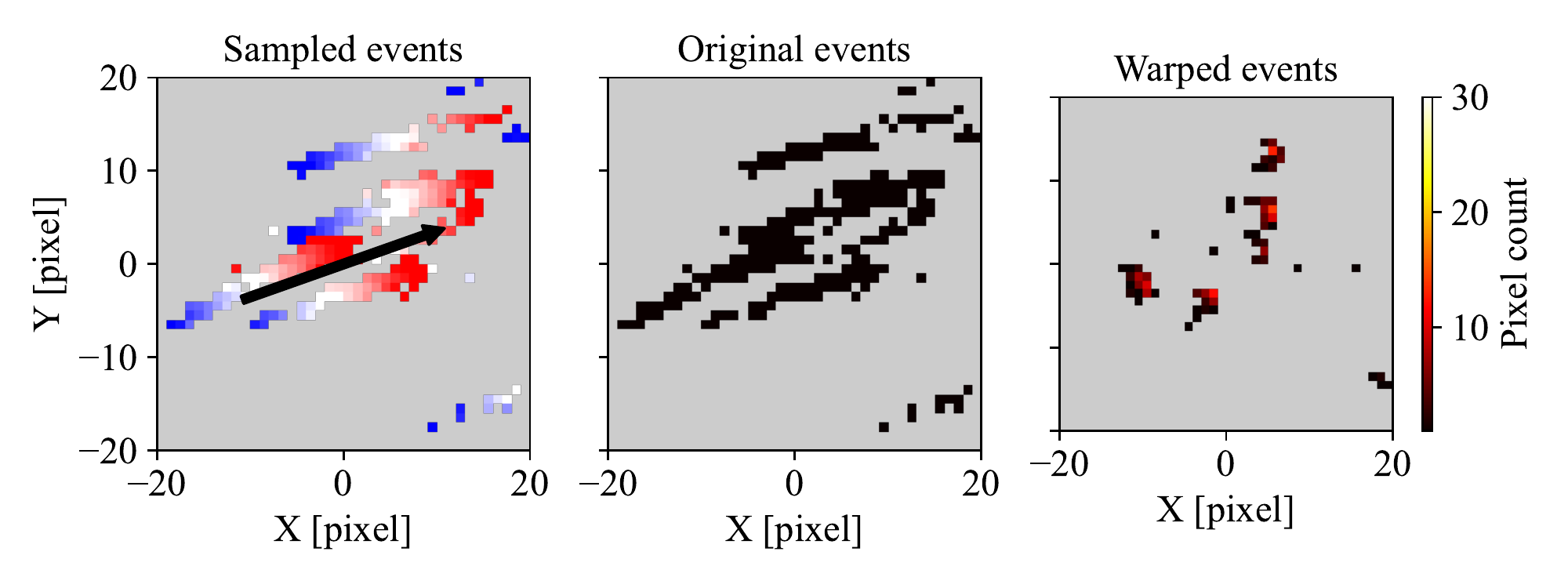}
    \includegraphics[width=0.29\columnwidth]{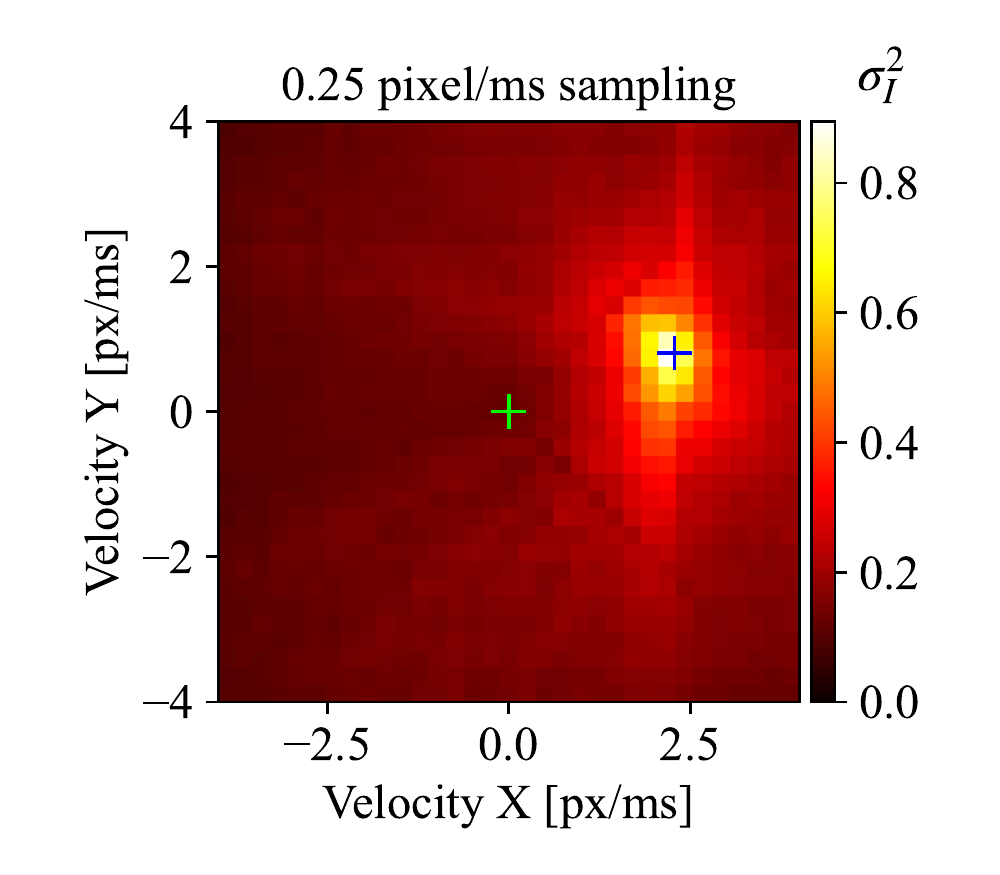}\\
    \includegraphics[width=0.70\columnwidth]{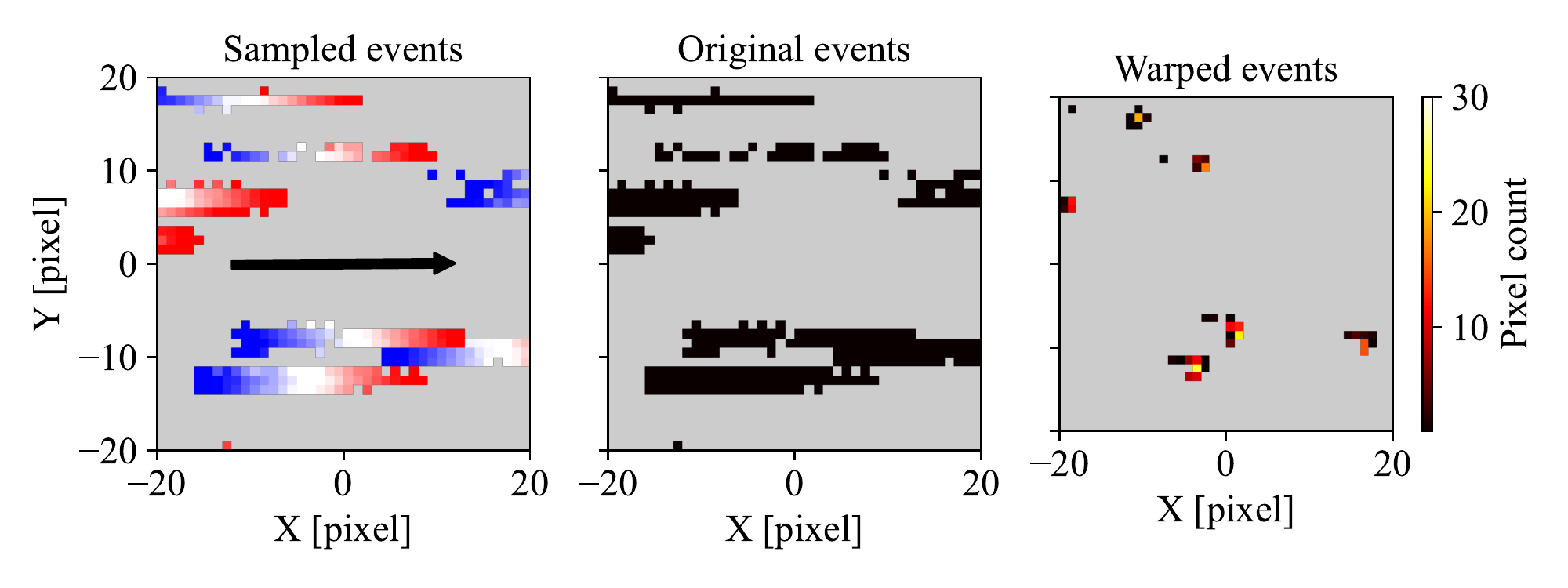}
    \includegraphics[width=0.29\columnwidth]{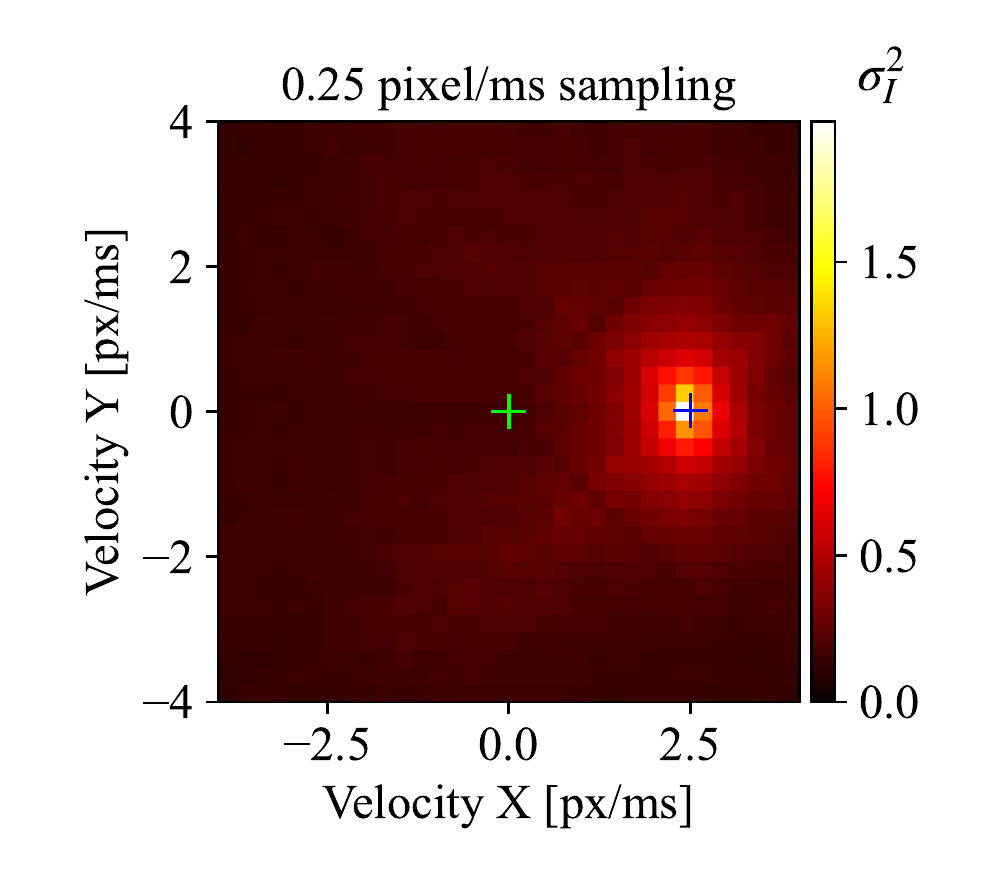}
    \caption{Sampled events from Figure~\protect\ref{fig:events_image} and corresponding images of warped events after applying motion compensation.
    Black arrow in leftmost sub-figures indicates the velocity estimate determined by finding the maximum
    in the 2-d variance distribution.}
    \label{fig:real_variance_processing}
\end{figure}

The two small squares in Figure~\ref{fig:events_image} indicate samples of 40$\times$40 pixels. The areas surrounding them are highlighted in Figure~\ref{fig:events_image_detail}.
Event tracks produced by the particles have a width of 2-5 pixels.
The procedure of the motion compensation algorithm is further demonstrated in Figure~\ref{fig:real_variance_processing}.
Here, the left-most sub-figure shows the sampled events color-coded with the time of arrival. For better visibility the second column shows only the currently ``active" pixels in the sampled sub-volume.
By applying motion-compensation, these pixels are aligned such that a maximum of events coincide, as shown in the third column of sub-figures.
In order to achieve this alignment, a measure based on the variance with respect to the space-time warped event data is used. The right column of Figure~\ref{fig:real_variance_processing} shows a distribution of the intensity variance spanned by a range of possible ``pixel velocities".
These variance intensity maps exhibit clearly detectable peaks that coincide with the most probable velocity estimate. In effect, this is very similar to the 2d cross-correlation map used in conventional PIV processing \cite{PIVBookRaffel:2018}.

\begin{figure}[htb]
    \centering
    \includegraphics[width=0.90\columnwidth]{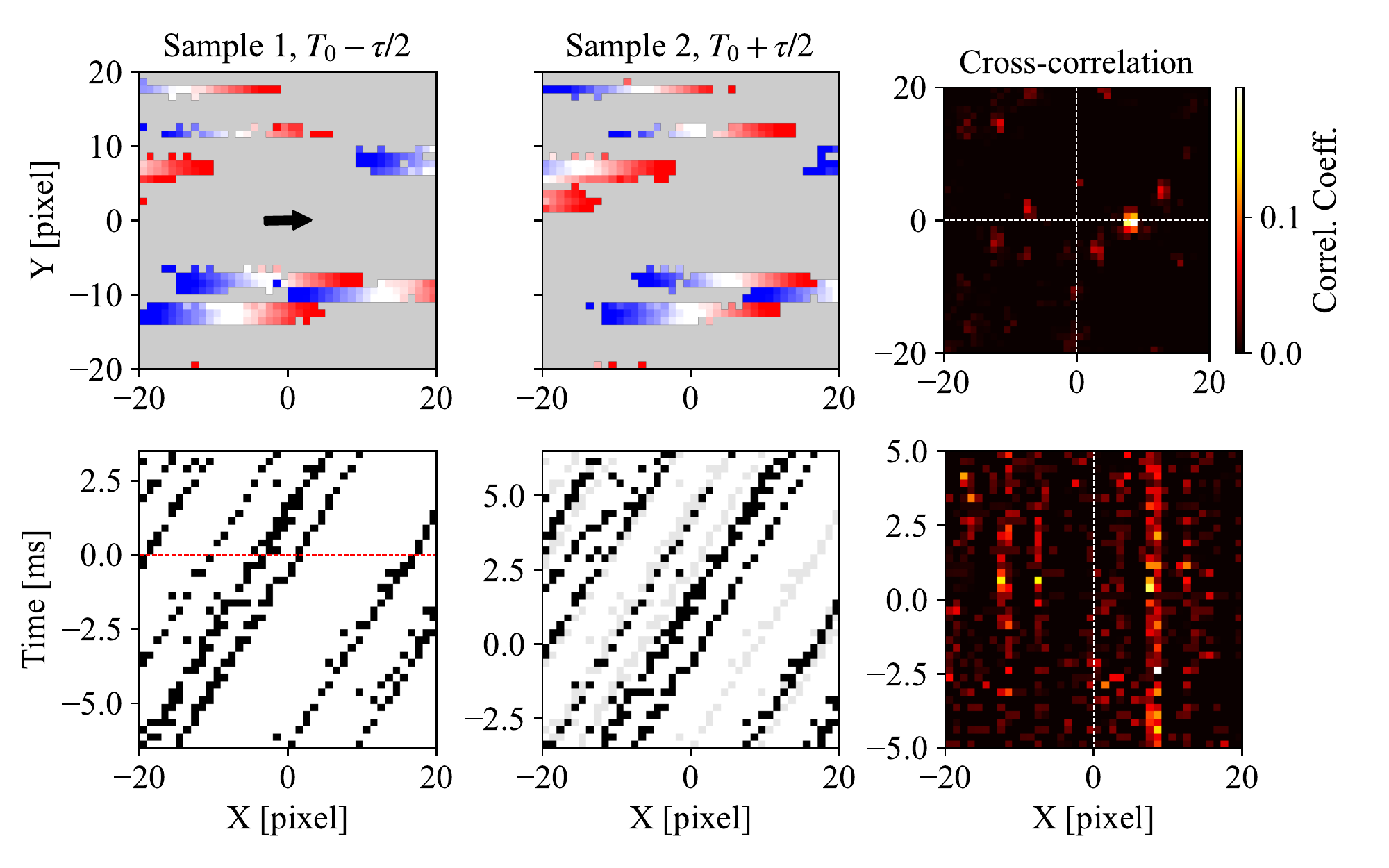}
    \caption{Top left, middle: Sampled events from Figure~\protect\ref{fig:events_image} obtained with time separation of $\tau = 3$\,ms (color map as in Figure~\protect\ref{fig:events_image_detail}).
    Bottom left, middle: $x-t$ projection of both samples.
    Bottom right: $x-t$ view of $N_T = 40$ individual cross-correlations over 10\,ms sample interval.
    Top right: cross-correlation map obtained by summing correlation planes below.
    Black arrow in upper left sub-figure indicates the velocity estimate determined by finding the maximum
    in the 2-d cross-correlation map.}
    \label{fig:real_corrsum_processing}
\end{figure}

Figure~\ref{fig:real_corrsum_processing} illustrates the sum-of-correlation approach on the same sample as in Figure~\ref{fig:real_variance_processing}.
Here, the final correlation plane is essentially equivalent to that of conventional 2D-2C PIV processing, sharing typical characteristics such as multiple correlation peaks arising through the correlation of events of one particle of the event of a different particles.
The lower right sub-figure in Figure~\ref{fig:real_corrsum_processing} also shows one of the main advantages of the sum-of-correlation method: any time-slices without events produce no correlation signal and have no contribution to the final correlation sum. Also noise in the individual correlation slices is averaged out.

% trim=left bottom right top

\begin{figure}[tb]
    \includegraphics[width=0.46\columnwidth]{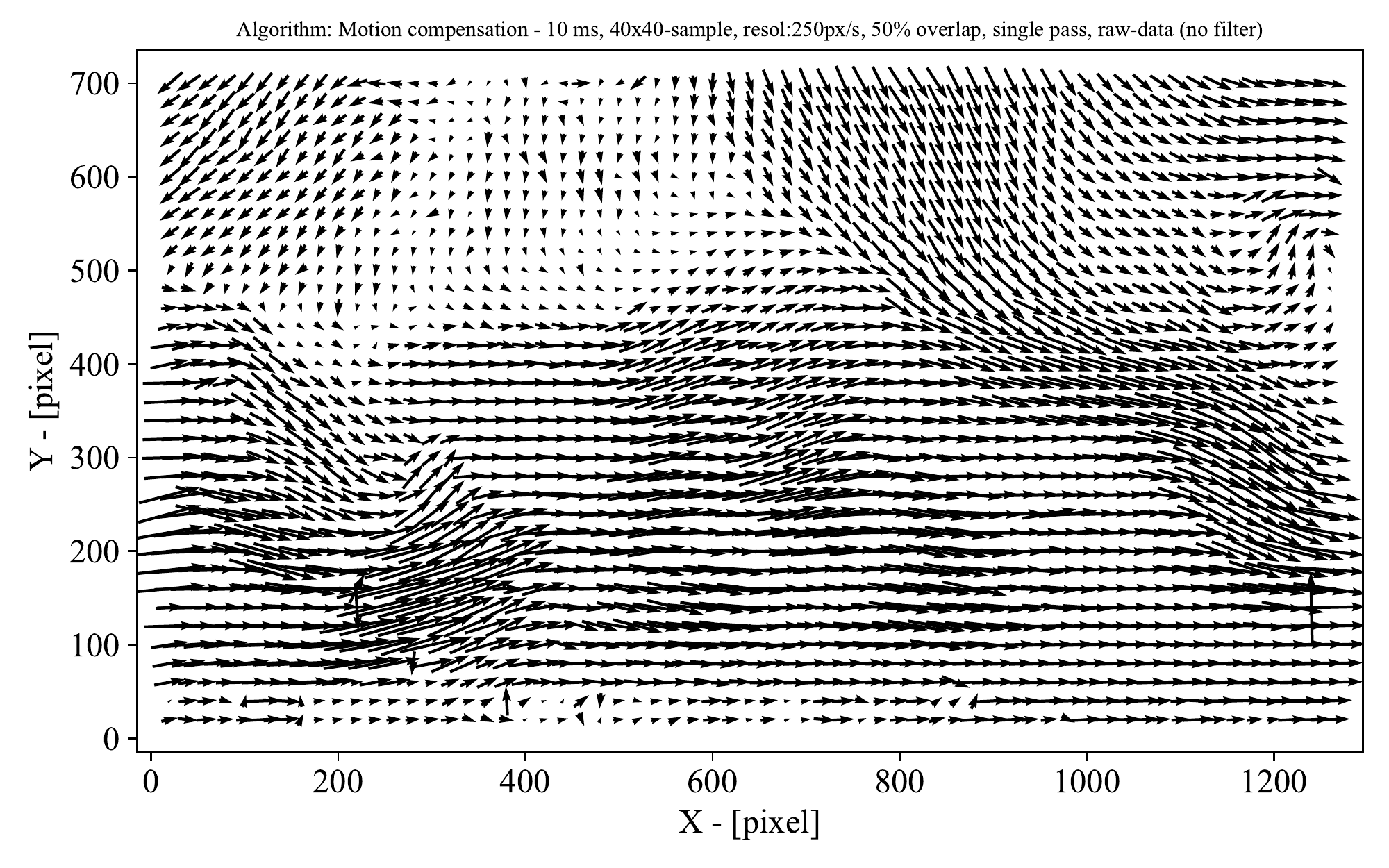}
    \includegraphics[width=0.53\columnwidth]{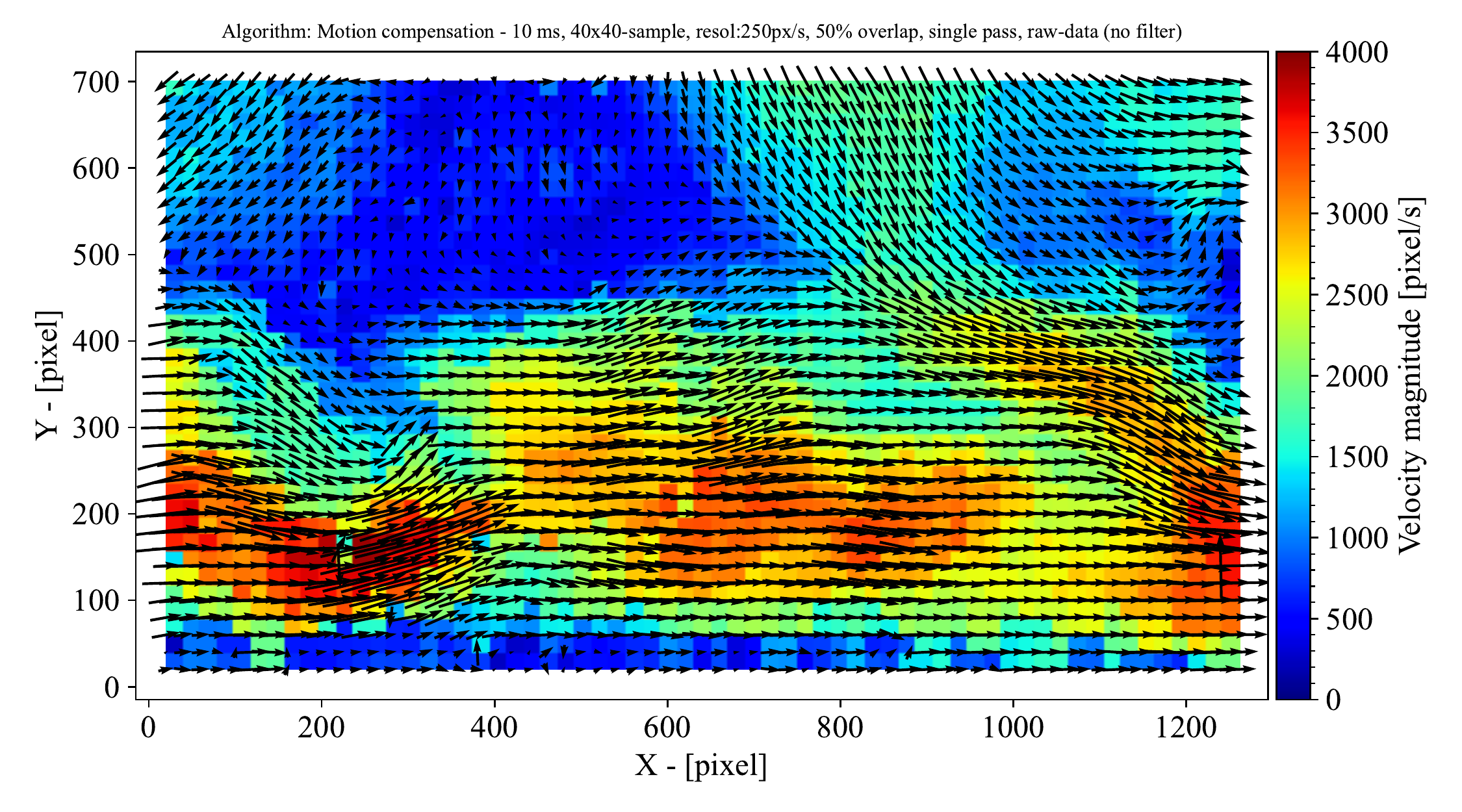}\\
    \includegraphics[width=0.46\columnwidth]{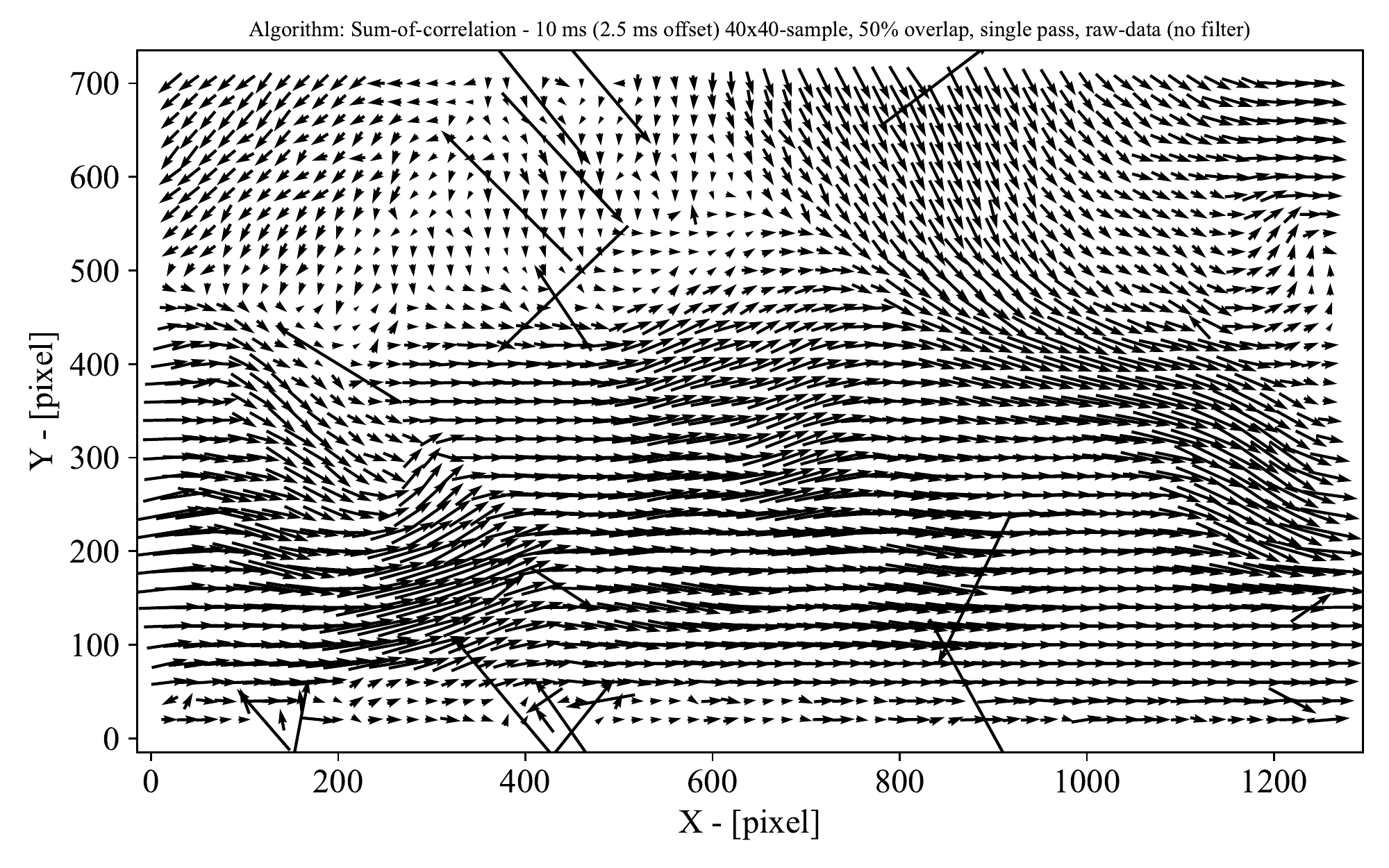}
    \includegraphics[width=0.53\columnwidth]{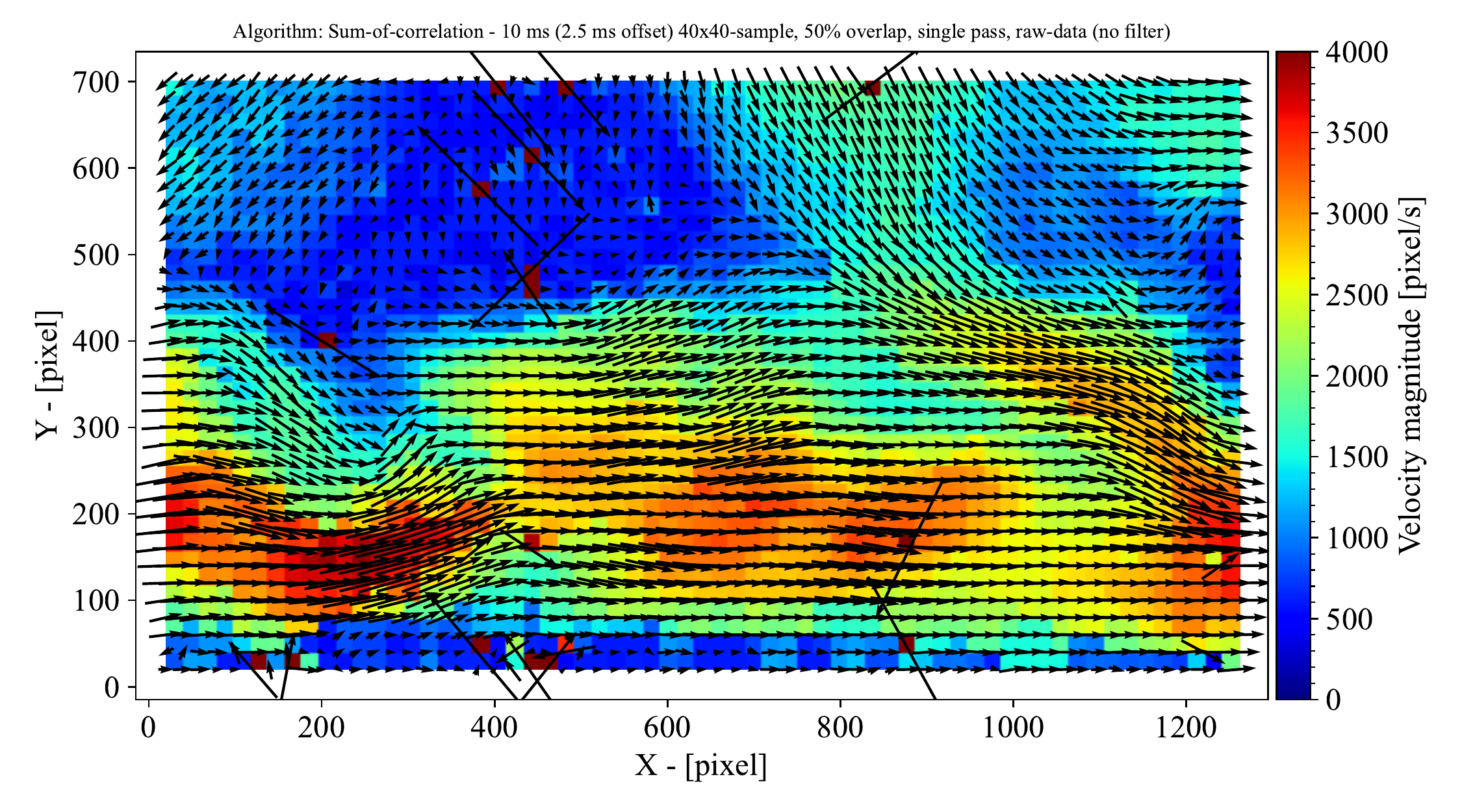}\\
    \includegraphics[width=0.46\columnwidth]{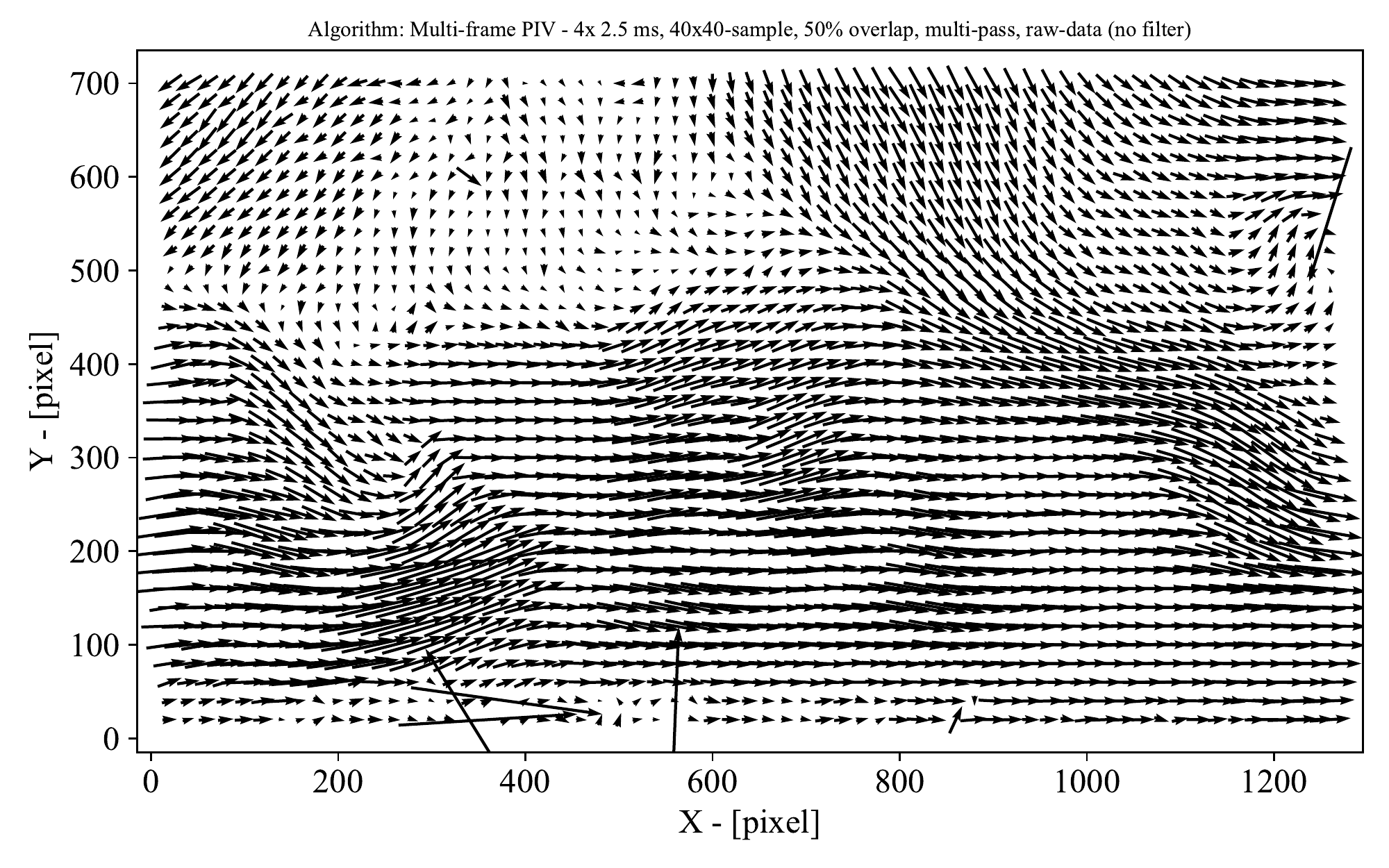}
    \includegraphics[width=0.53\columnwidth]{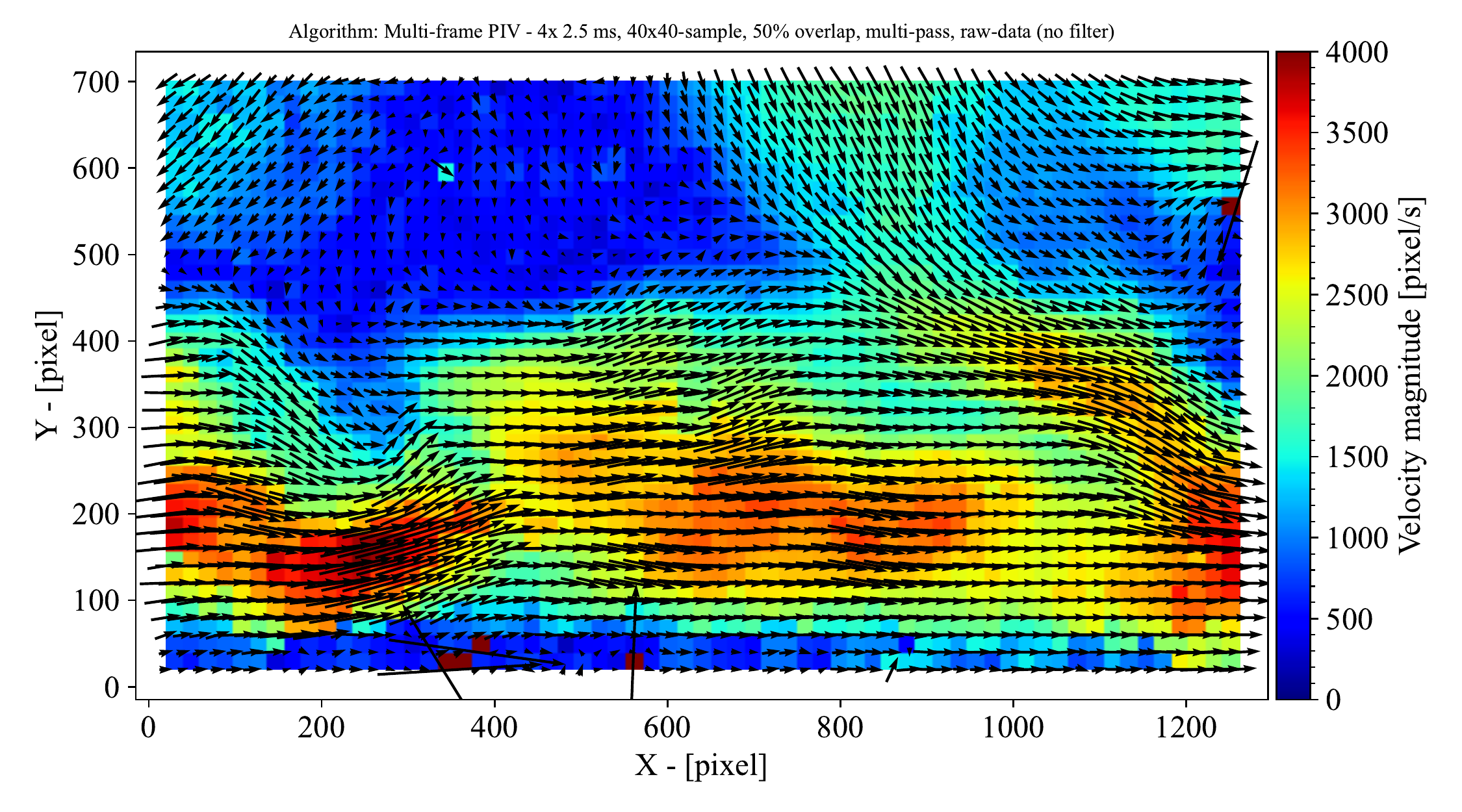}
    \caption{Velocity fields obtained with the proposed processing algorithms using event data shown in Figure~\protect\ref{fig:events_image}.
    Top row: motion compensation approach, middle row: sum-of-correlation approach, bottom row: 4-frame PIV analysis using images created from events of duration 1.5\,ms.
    %Both are single-pass implementations without grid-refinement or validation applied.
    Physical velocity is obtained by multiplying with the scaling factor, $m = 0.021$\,mm/pixel.
    Animations of the data obtained with the sum-of-correlations approach are provided in the supplementary material.}
    \label{fig:eval_onepass}
\end{figure}

In PIV processing, the displacement should be limited to about one fourth the sample dimension to warrant an optimal displacement recovery by restricting the in-plane loss of particle image pairs (so called ``one-quarter rule", see e.g. \cite{PIVbookAdrianWesterweel,PIVBookRaffel:2018}).
On first sight, this sort of restriction is not present for the described variance based motion compensation algorithm. This can be observed in Figure~\ref{fig:real_variance_processing}, bottom, which shows a displacement of about 25 pixels in a sample of 40$\times$40 pixel.
The event data shown in Figure~\ref{fig:events_image} contains particle image tracks whose length exceeds 30 pixel (cf. Figure~\ref{fig:eval_onepass}).
Nonetheless, fast moving particles will produce a reduced number of events in the sampled space-time sub-volume which in turn reduces the reliability of the velocity estimate.
The sum-of-correlation approach, on the other hand, should adhere to the ``one-quarter rule" in order to reliably line up the event tracks from both volumes. Faster moving particle images are captured by reducing the separation $\tau$  or by increasing the size of the sample.

As an example, the result of a single-pass processing of the water flow is provided in Figure~\ref{fig:eval_onepass} using a fixed sample size of 40$\times$40 pixels on a time-slab of 10\,ms.
For the motion compensation algorithm the velocity search domain is set to $v_x = \left[ -1,+5\right ]$\,pixel/ms and $v_y = \left[ -2,+2\right ]$\,pixel/ms at an initial sampling resolution of 0.250\,pixel/ms which is iteratively refined on a 5$\times$5 neighborhood around the maximum variance until the resolution is below a cut-off value of 0.05\,pixel/ms.
The sum-of-correlation approach uses an temporal offset between the samples of $\tau$ = 2.5\,ms (1/4th of the sample interval). Each pair of $x-y-t$ sub-volumes is resampled to $N_T$ = 20 planes from which 20 individual correlation planes are calculated and subsequently summed.
Finally, Figure~\ref{fig:eval_onepass} presents the velocity field obtained using a conventional PIV processing algorithm.
This is realized by first generating classical PIV recordings from the event data simply by combining events from a given time interval (1.5\,ms) into a single image frame using the time stamps as intensity values. The temporal spacing between the image frames is chosen at $\Delta t$ = 2.5\,ms which equates to a frame rate of 400\,Hz.
No further image processing, such as filtering or intensity thresholding, is performed prior to applying an iterative, coarse-to-fine cross-correlation scheme using 4 frames per time-step \cite{LynchScarano:2013}.

Overall, the agreement between the results of the three processing approaches is very good; the differences are on the order of 100\,pixel/s.
The data obtained with the motion compensation scheme appears noisier in comparison to the other two approaches but exhibits the least amount of spurious (outlier) vectors.
The sum-of-correlation scheme has difficulty in the quiescent regions whereas the conventional PIV processing exhibits most of the outliers near the wall ($y \approx $ 40 pixel).
Due to their reduced motion, particles in the quiescent regions produce a reduced number of events such that the algorithms fail to produce reliable results in these areas.

The processing speed of the algorithms depends on numerous parameters and varies linearly with the number of sampling points and sub-sample size in space and time. For the present data, the sum-of-correlation approach requires about 1~s per data per time-step (single thread processing on a 4 GHz Intel CPU) and is about five times faster than the motion compensation approach.

\section{Performance assessment}
\label{sec:performance}
While the previously shown measurements of the water flow demonstrate the feasibility of EBIV, little can be learned regarding measurement accuracy and uncertainty.
One approach to at least assess the influencing factors is to use an experiment or simulation with known characteristics.
This can be either be achieved using synthetic data, which is a challenge in itself (see e.g. \cite{HuDelbruck:2021}), or by recording event data from a clearly defined flow.
In the present case, the ``flow" is produced by a solid body rotation of constant angular speed achieved by placing a plate of fixed particles on a turn table (i.e. record player). Sufficiently small particles are provided by reflective spray containing small glass spheres of about 50$\mu$m diameter. The lens of the event camera is fitted with a ring of white-light LEDs such that the illumination is essentially on-axis with the lens.
Event data is acquired for the inner radius extending out to $R \approx$\,60\,mm and the outer radius in the range $85 < R < 150$\,mm with the turn table operating at 45\,rpm.

\begin{figure}[tb]
\centering
    \includegraphics[width=0.49\columnwidth]{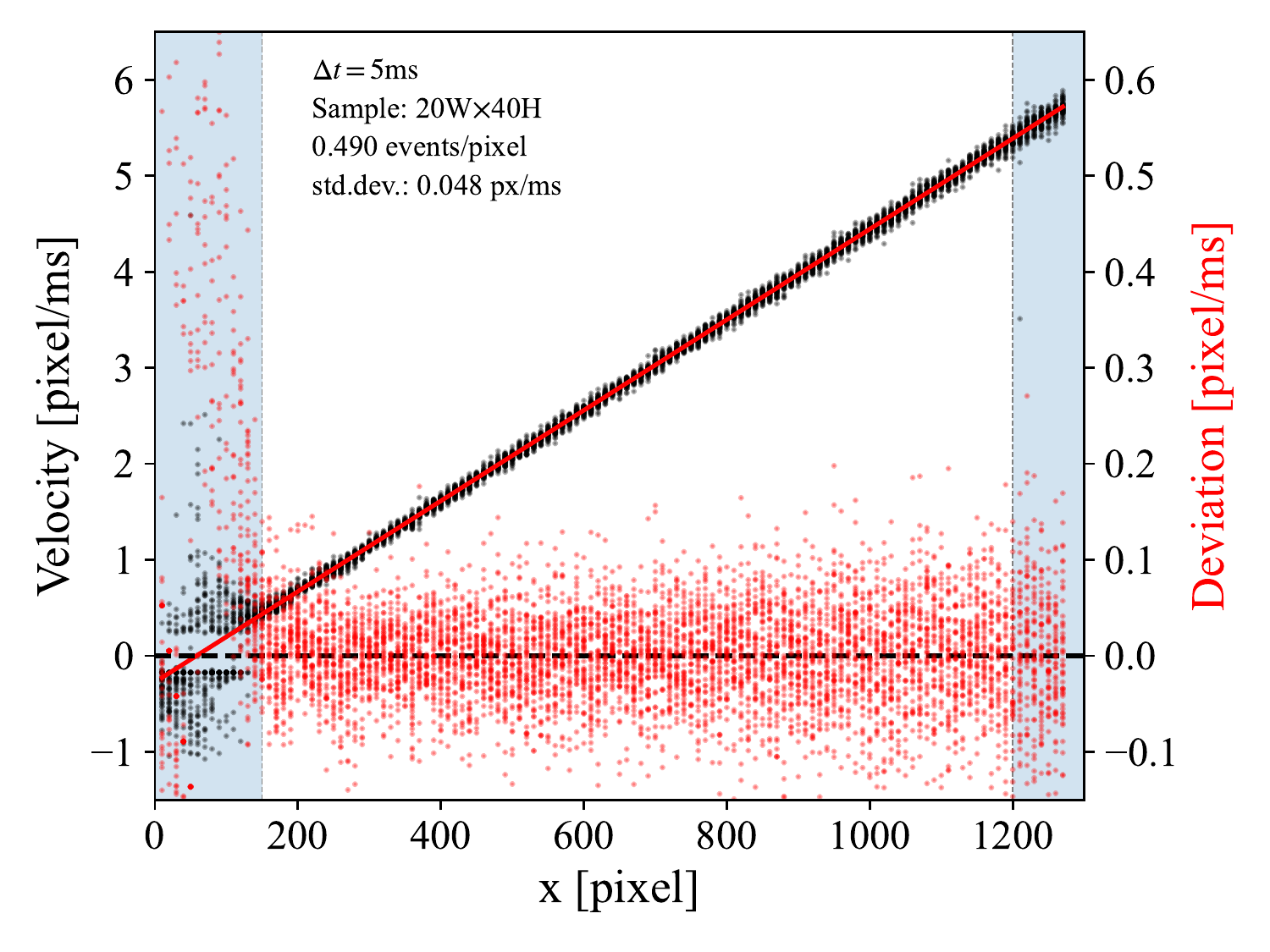}
    \includegraphics[width=0.49\columnwidth]{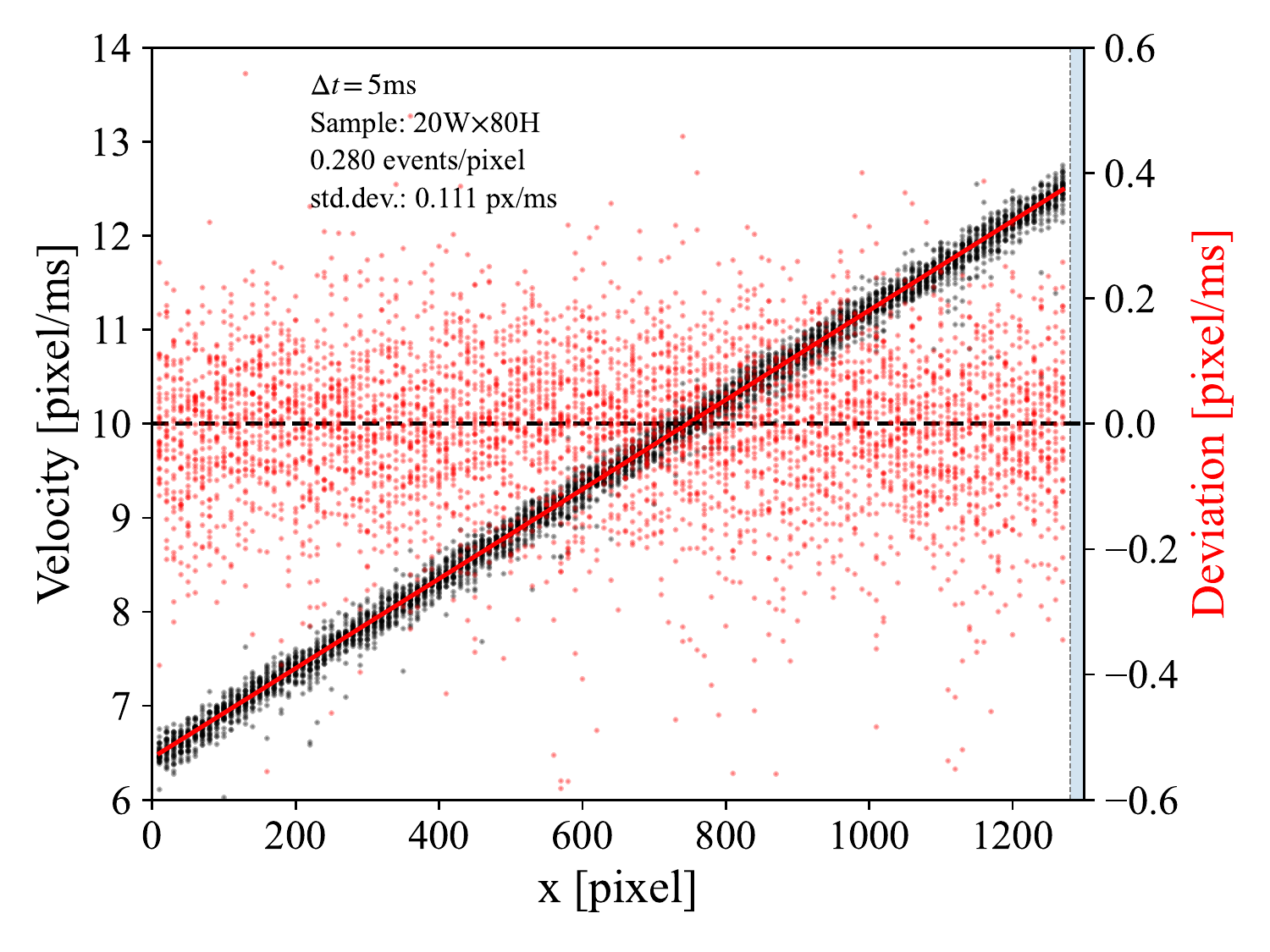}\\
    \includegraphics[width=0.49\columnwidth]{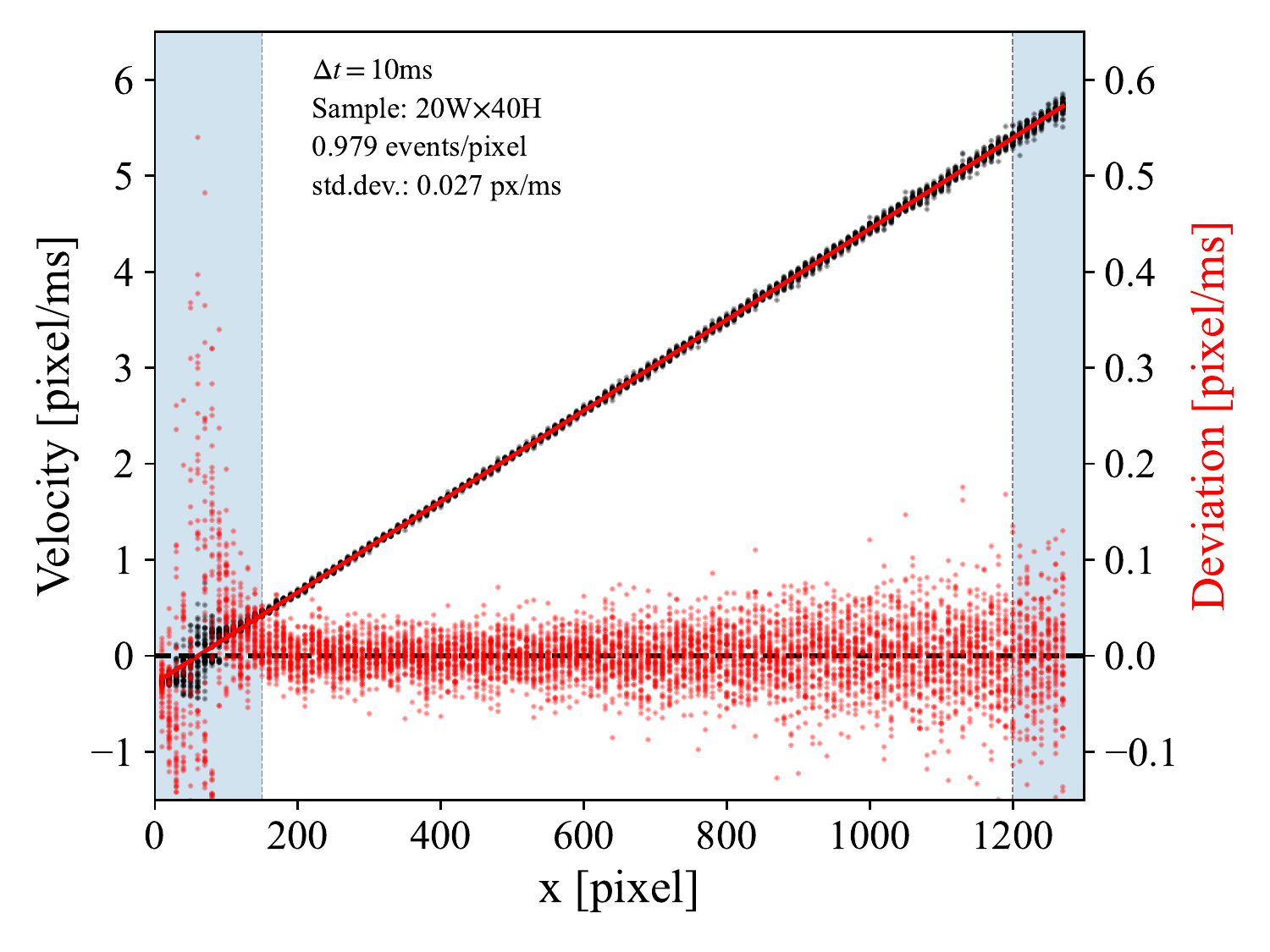}
    \includegraphics[width=0.49\columnwidth]{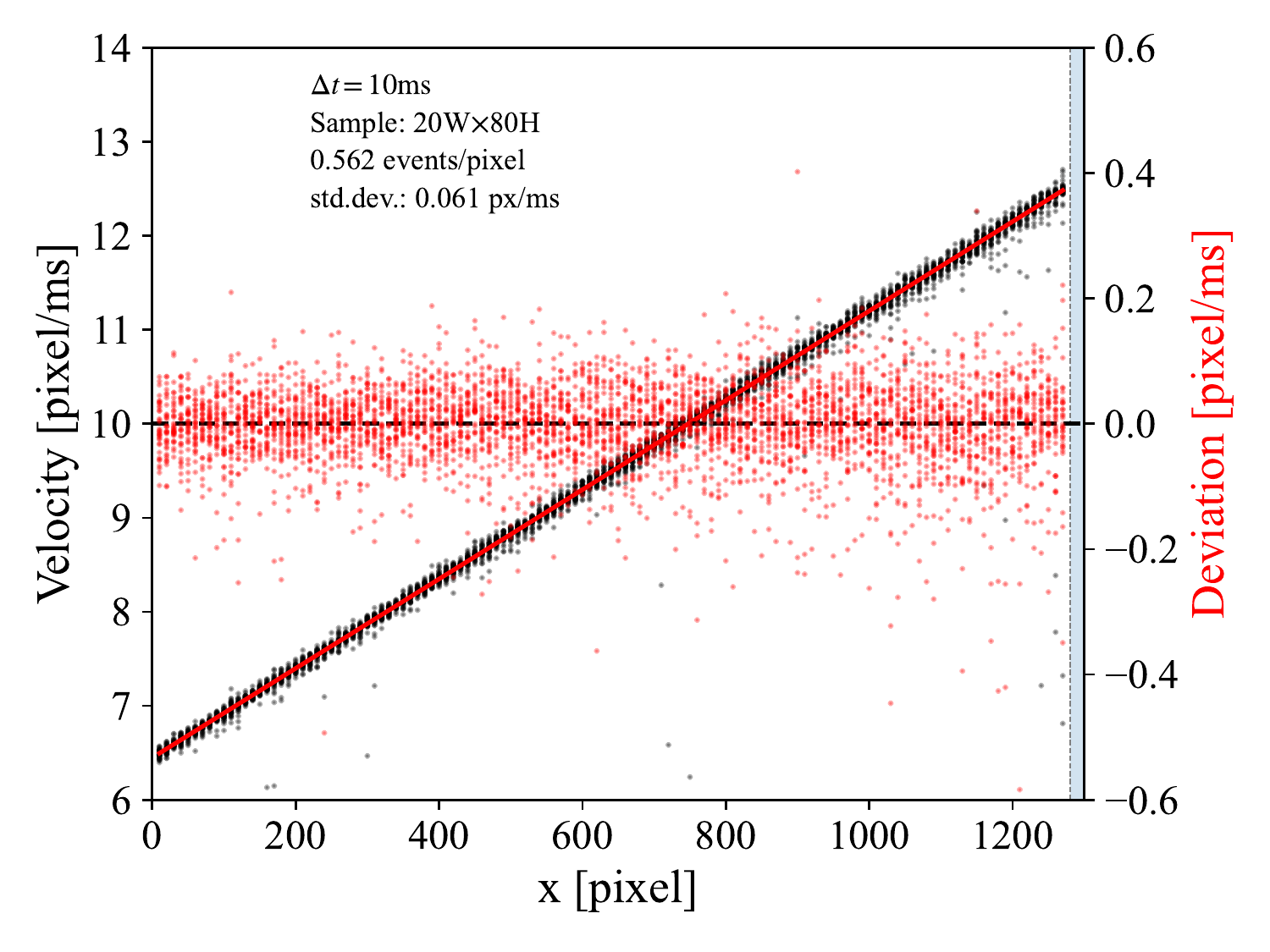}\\
    \includegraphics[width=0.49\columnwidth]{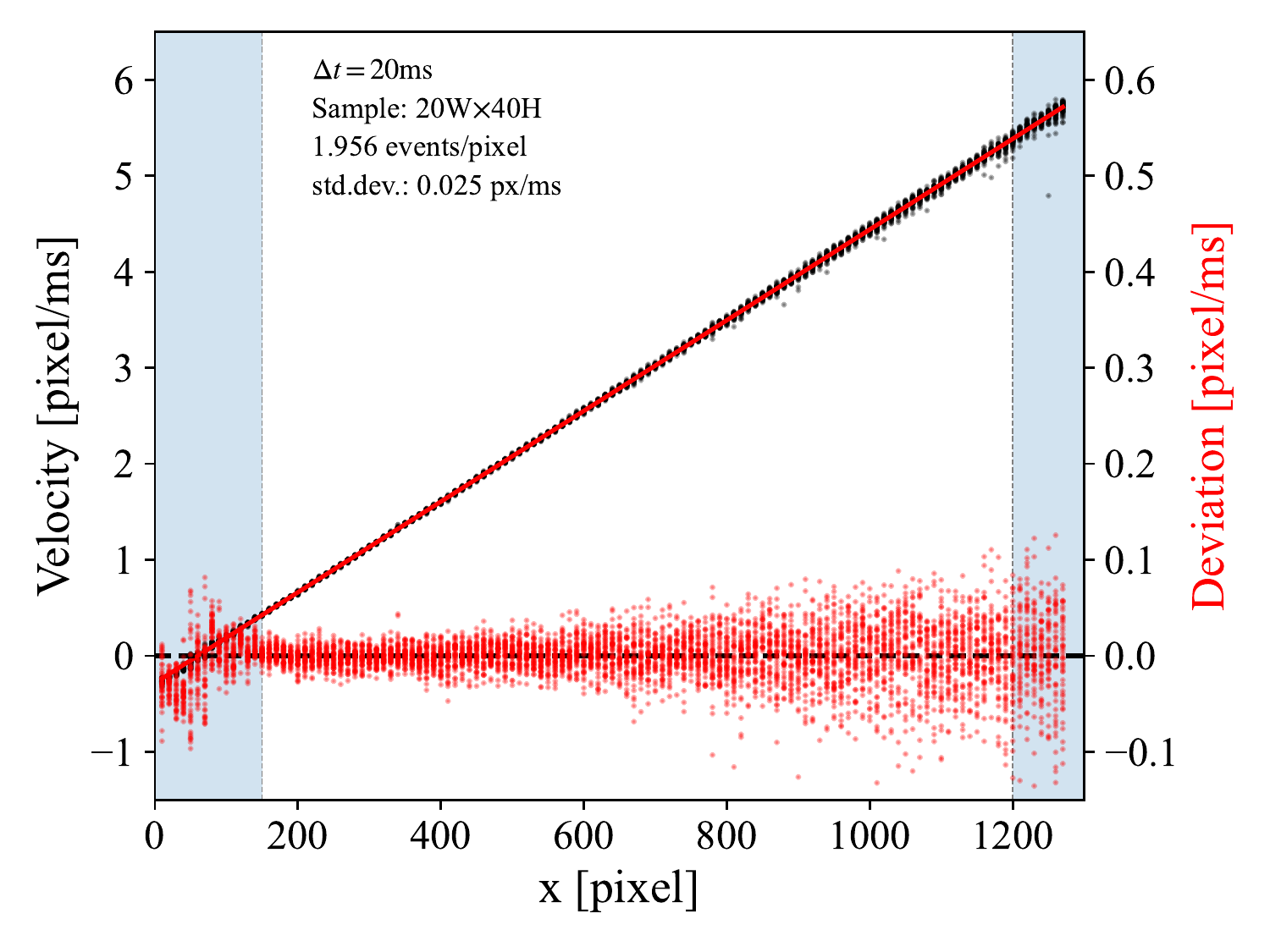}
    \includegraphics[width=0.49\columnwidth]{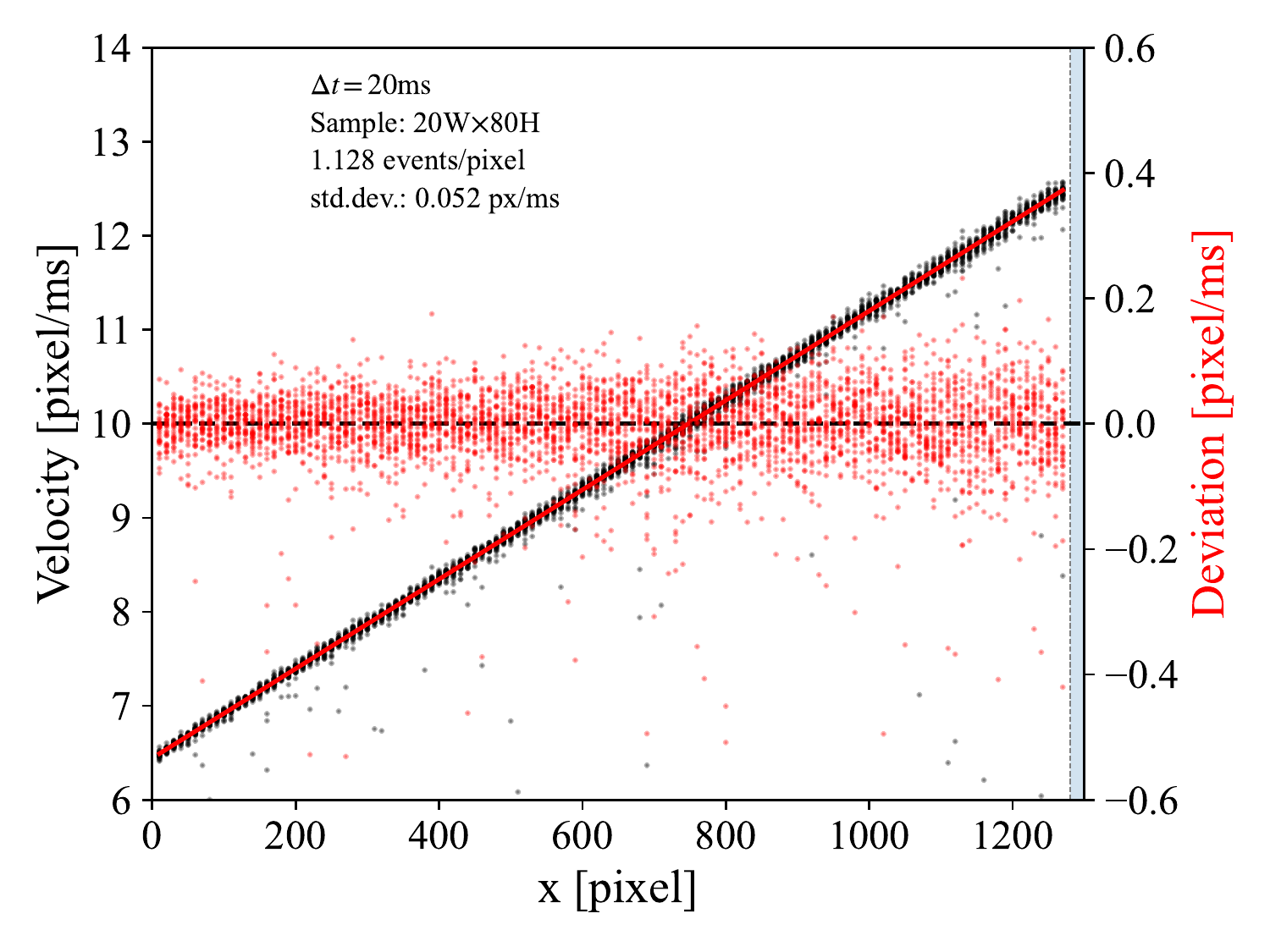}%
    \caption{Solid-body rotation data obtained by imaging small particles on a turn table rotating at 45\,rpm.
    The same data set is processed using different event samples times: 5, 10 and 20\,ms. Left column is processed with samples of 20$\times$40 pixel, the right with 20$\times$80 pixel. Data in the grayed out region are excluded.}
    \label{fig:solidbody_data}
\end{figure}

Acquired event data is processed along the radius using the previously described motion compensation algorithm with a fixed sample size of 20(W)$\times$40(H) pixels for the inner radii and increased to 20(W)$\times$80(H) pixels for $R > 85$\,mm to capture a sufficient number of events.
Velocity information is determined by sampling the 2 second long records with sampling intervals $T$ of 5, 10 and 20\,ms.
Figure~\ref{fig:solidbody_data} presents both the raw velocity estimates in the form of scatter plots along with a linear fit to the data.
In agreement to the underlying the solid body rotation, the velocity increases linearly with increasing radius.
The deviation of the data from the linear fit is plotted with red dots and exhibits a normal distribution.
The scatter increases with increasing radius and decreases with increasing sampling time $T$.
The measurement of the solid body rotation clearly shows that measurements up to (and beyond) 12\,pixel/ms (or 12\,000 pixel/s) are feasible and show no systematic deviation.

In effect, the sampling time defines how many events of a particle are captured and used for the estimation of the velocity.
Unlike PIV, in-plane loss of particle images is not really relevant here because other particles come into view and also contribute to the velocity estimate.
For the 10\,ms sampling condition the dynamic velocity range (DVR, \cite{Adrian:1997}) can be estimated at (3.5/0.020) = (175$\div$1) which is comparable to that of 2d-2c PIV with sampling at 32$\times$32\,pixel.
While the rms deviation reduces by 50\% when doubling the sampling time from 5\,ms to 10\,ms, the reduction is not as significant when doubling the sampling time from 10\,ms to 20\,ms.
Effects such as particle path curvature, particle acceleration/deceleration, non-linearities in the event formation, etc., will contribute to the velocity uncertainty.
It is clear, that additional Monte-Carlo simulations and modelling of the EBV imaging process (which can be described as a Poisson distribution) are required to further quantify the factors influencing measurement uncertainty.
While not presented here, it should be noted that the sum-of-correlation approach is not able to capture the entire range of radii without adjusting the temporal sample offset $\tau$ to warrant that the correlation peak is captured at higher velocities.

\begin{figure}[htb]
\centering
    \includegraphics[width=0.45\columnwidth]{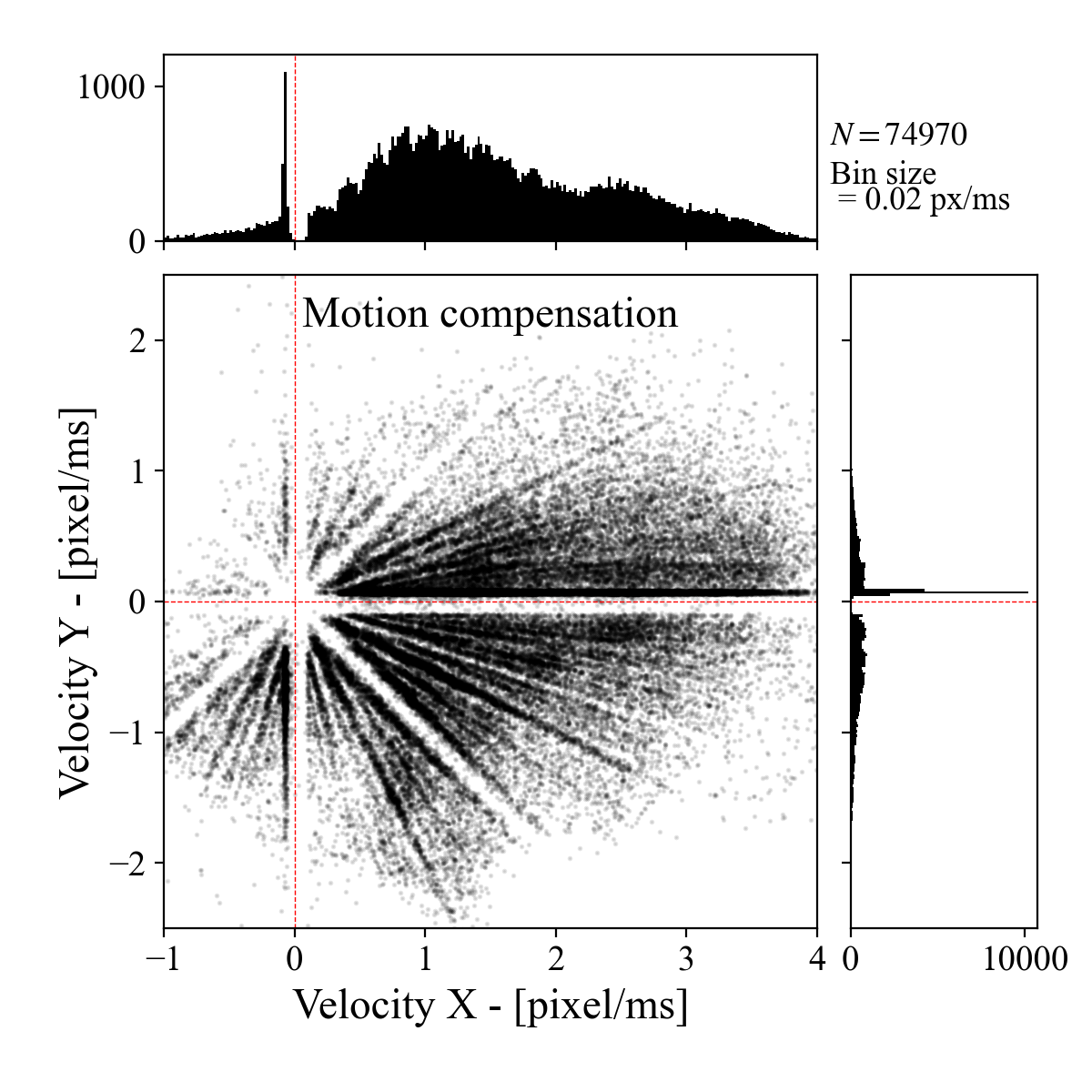}
    \includegraphics[width=0.45\columnwidth]{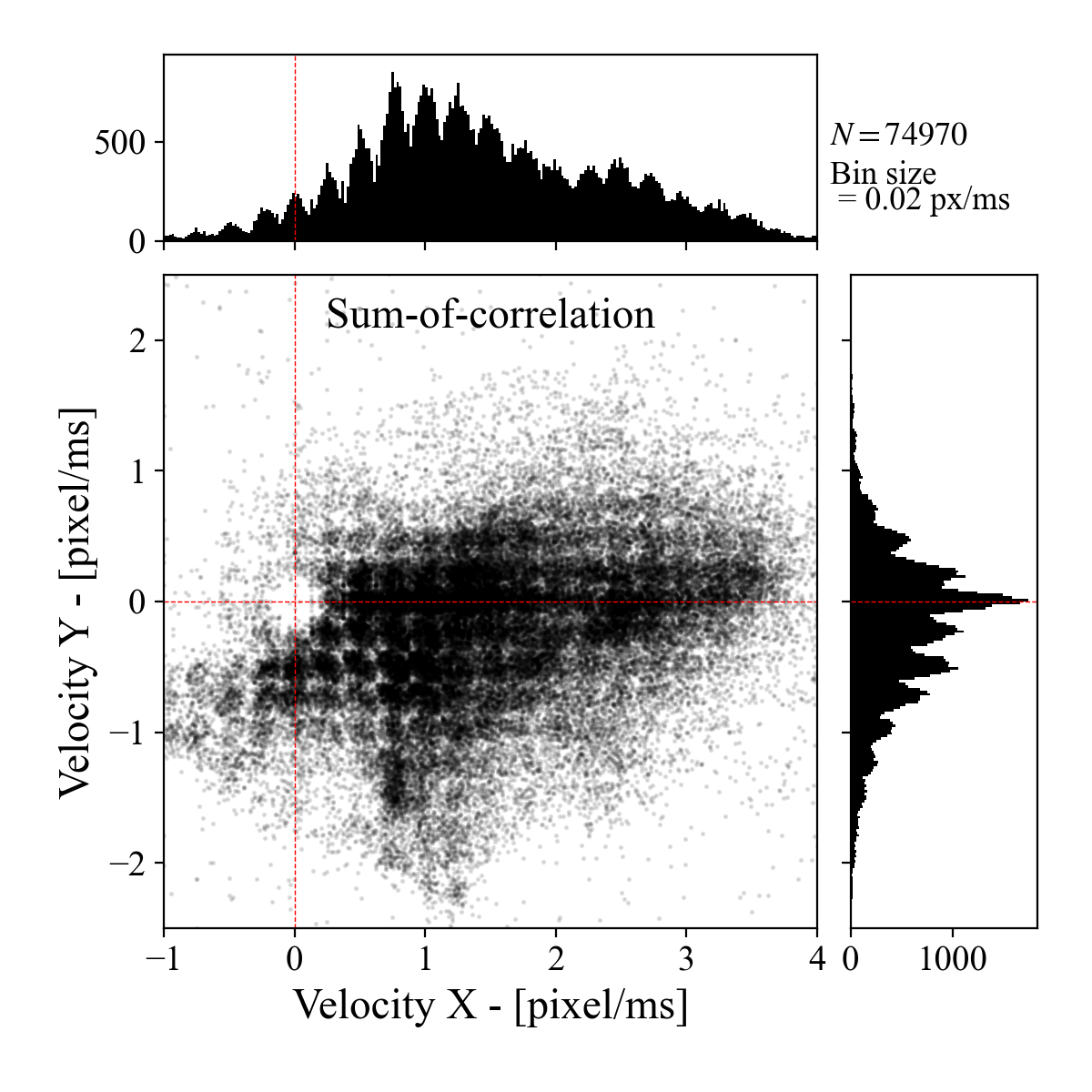}\\
    \includegraphics[width=0.45\columnwidth]{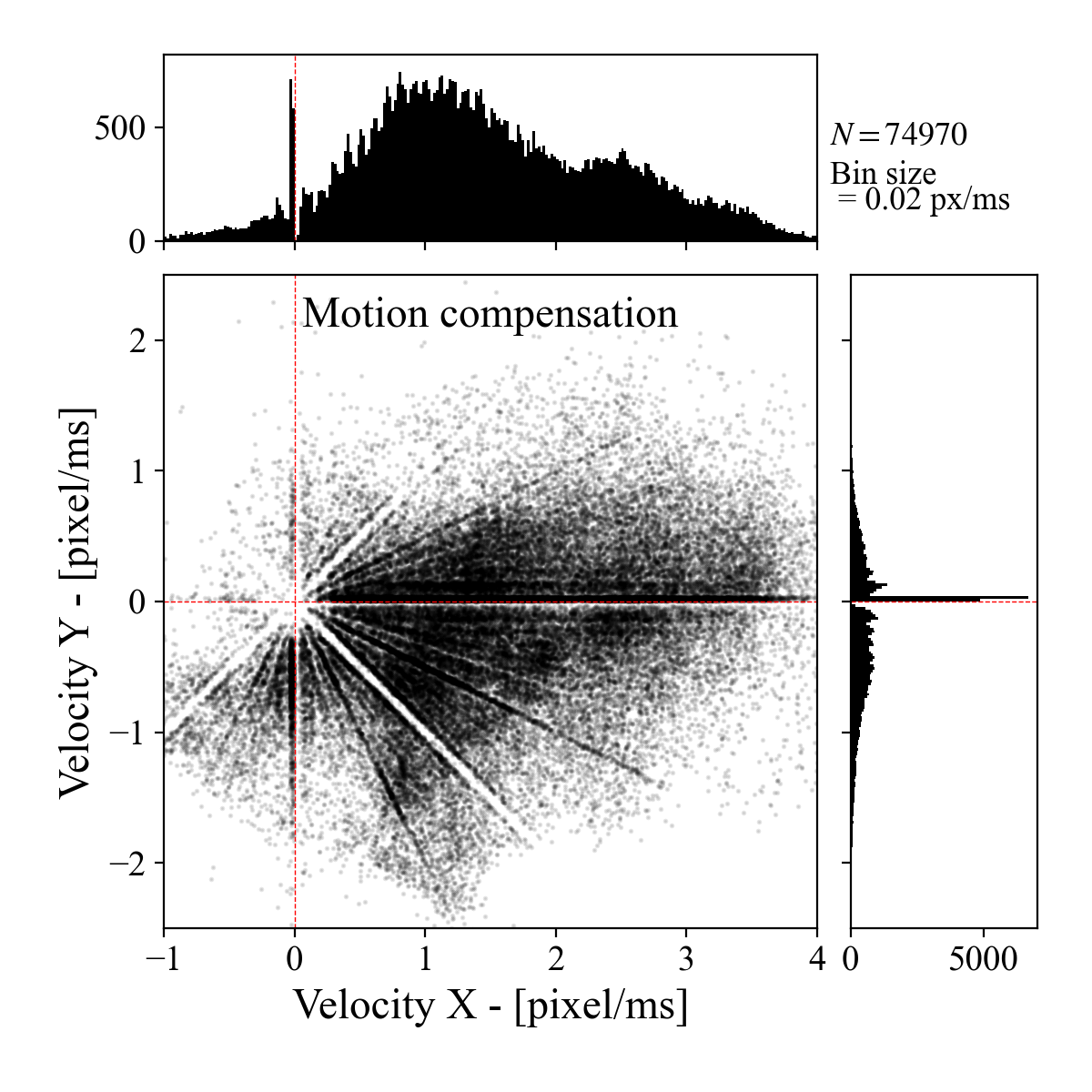}
    \includegraphics[width=0.45\columnwidth]{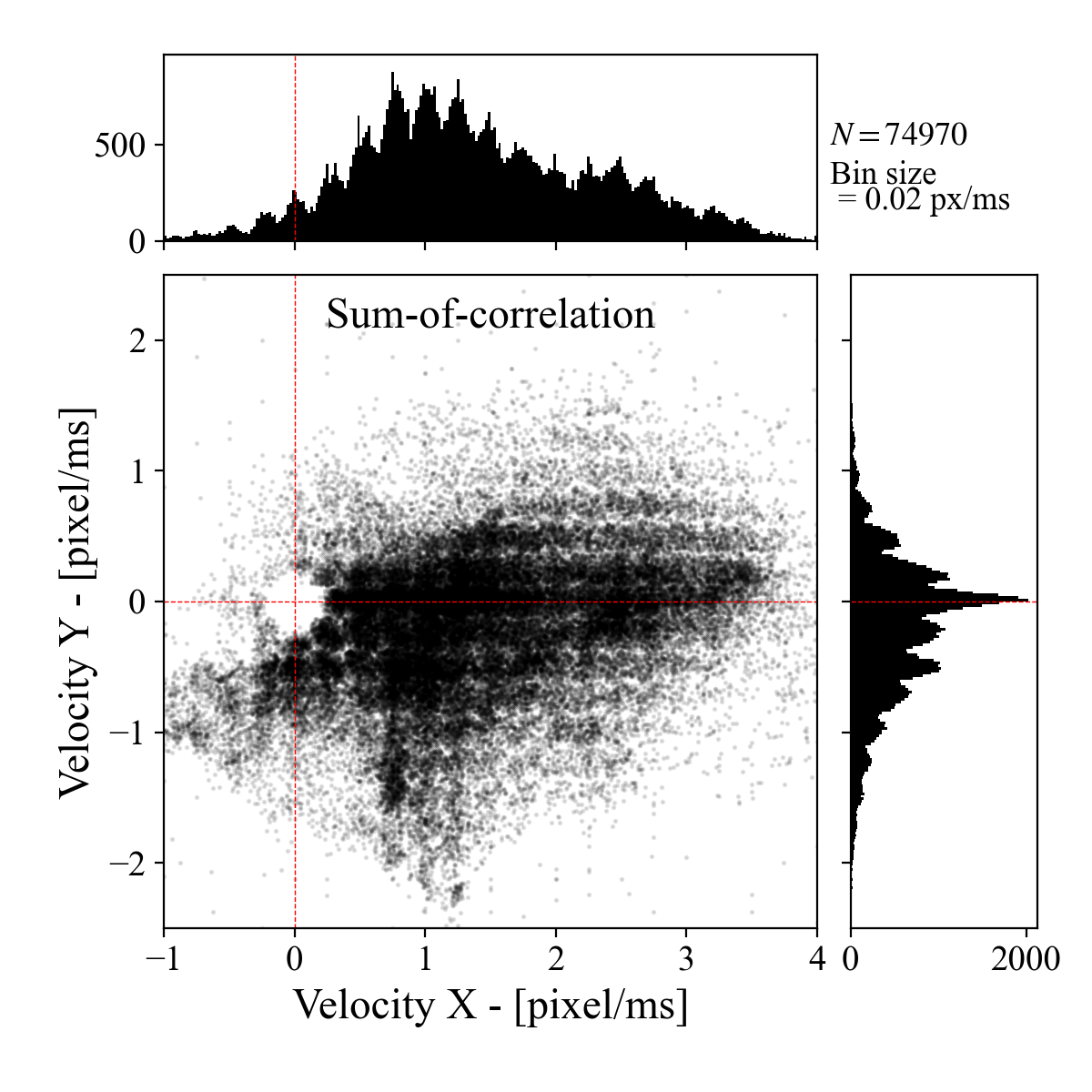}\\
    \includegraphics[width=0.45\columnwidth]{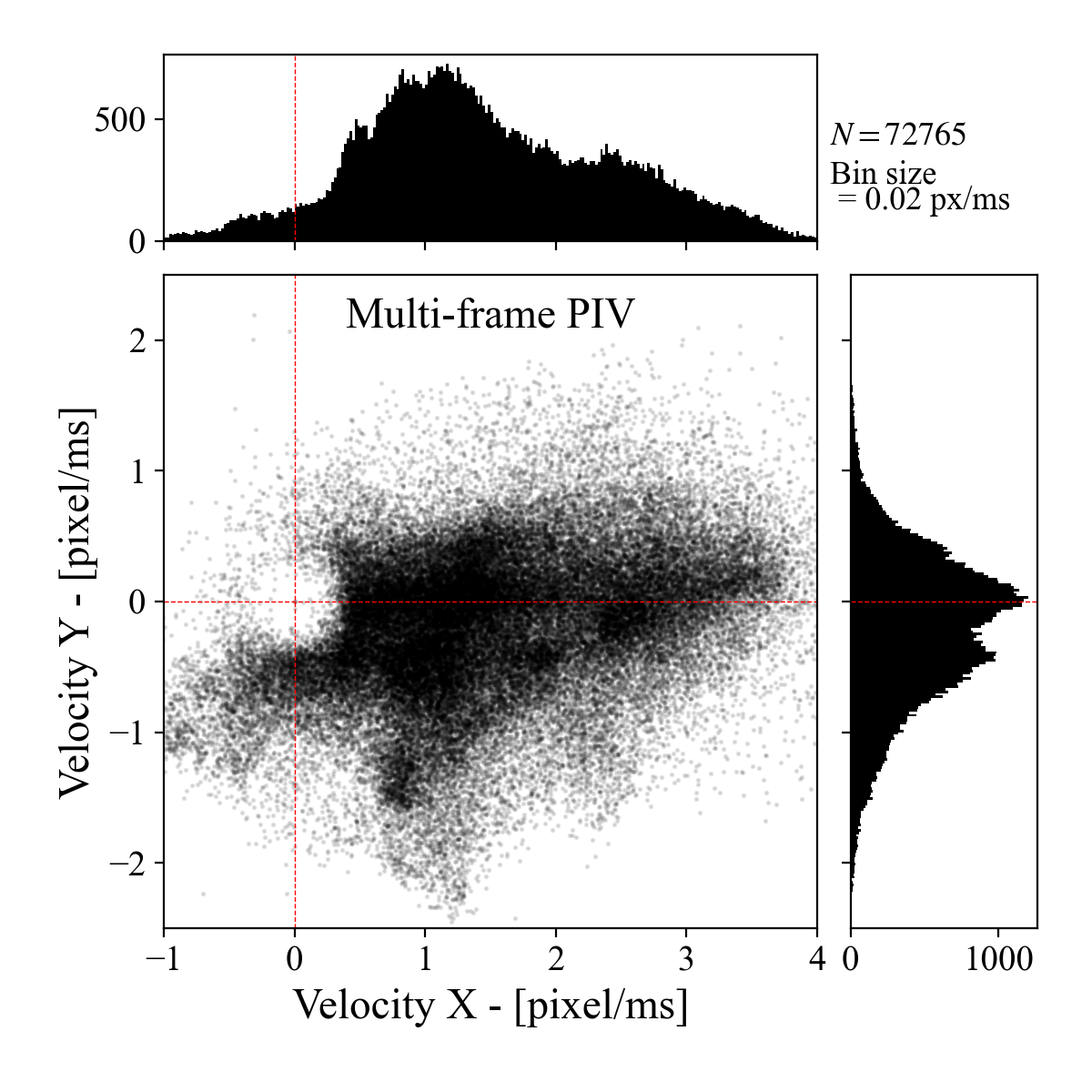}
    \caption{Histograms obtained from 35 reconstructed flow fields of the turbulent water flow using three different algorithms for time-intervals of $T$ = 10\,ms (top row) and $T$ = 20\,ms (middle row).
    Bottom plot shows result obtained using multi-frame PIV applied to images separated by 2.5\,ms, each containing 1.5\,ms of events.
    }
    \label{fig:histograms}
\end{figure}

Another common practice of assessing PIV data quality is to plot the histograms of the displacement data in order to uncover artifacts such as pixel locking.
Using 1s worth of processed event-data from one of the experiments (turbulent water flow, see Sec.~\ref{sec:waterflow-exp}) both 2d scatter plots and projected (1d) histograms of the velocity data are provided in Figure~\ref{fig:histograms}.
The first four sub-figures shows velocity scatter plots obtained with the motion compensation approach (left) and sum-of-correlation approach (right) for time samples of $T$ = 10\,ms (top row) and $T$ = 20\,ms (second row).
The motion compensation data results from an iterative scheme that reevaluates the area around the most probable velocity value is the variance map using a grid of 5$\times$5 samples. The cut-off is chosen at 0.025\,pixel/ms.
All sub-figures in Figure~\ref{fig:histograms} clearly shows effects similar to pixel-locking found in PIV.
In the present case, the imaged turbulent flow should show a uniform distribution; the gridded nature of velocity clustering is non-physical.
In particular, the motion compensation algorithm produces pronounced star-shaped features that can also be observed in the variance maps of Figure~\ref{fig:real_variance_processing}, right.
Unlike pixel locking in PIV, these clustering artifacts are believed to be systematic errors that are purely associated with the respective processing algorithms. This, however, does not imply that event-based imaging does not suffer other short-comings that propagate into the velocity estimates.
In order to further characterize the sources of uncertainty, the event-based imaging of small particles has to be modelled using, for instance, approaches such as put forth by Hu et al.~\cite{HuDelbruck:2021}.

Noteworthy in Figure~\ref{fig:histograms} is the absence of velocities near zero. This is due to the fact, that particles at rest do not produce sufficient events to be registered.
This is clearly a shortcoming of the measurement technique for which no obvious solution is presently available.
In principle by increasing the sampling time $T$ more events can be collected allowing velocities closer to zero to be captured.

\section{EBIV Measurement in Air}
\label{sec:airflow}
\begin{figure}[htb]
\centering
    \includegraphics[width=0.69\columnwidth]{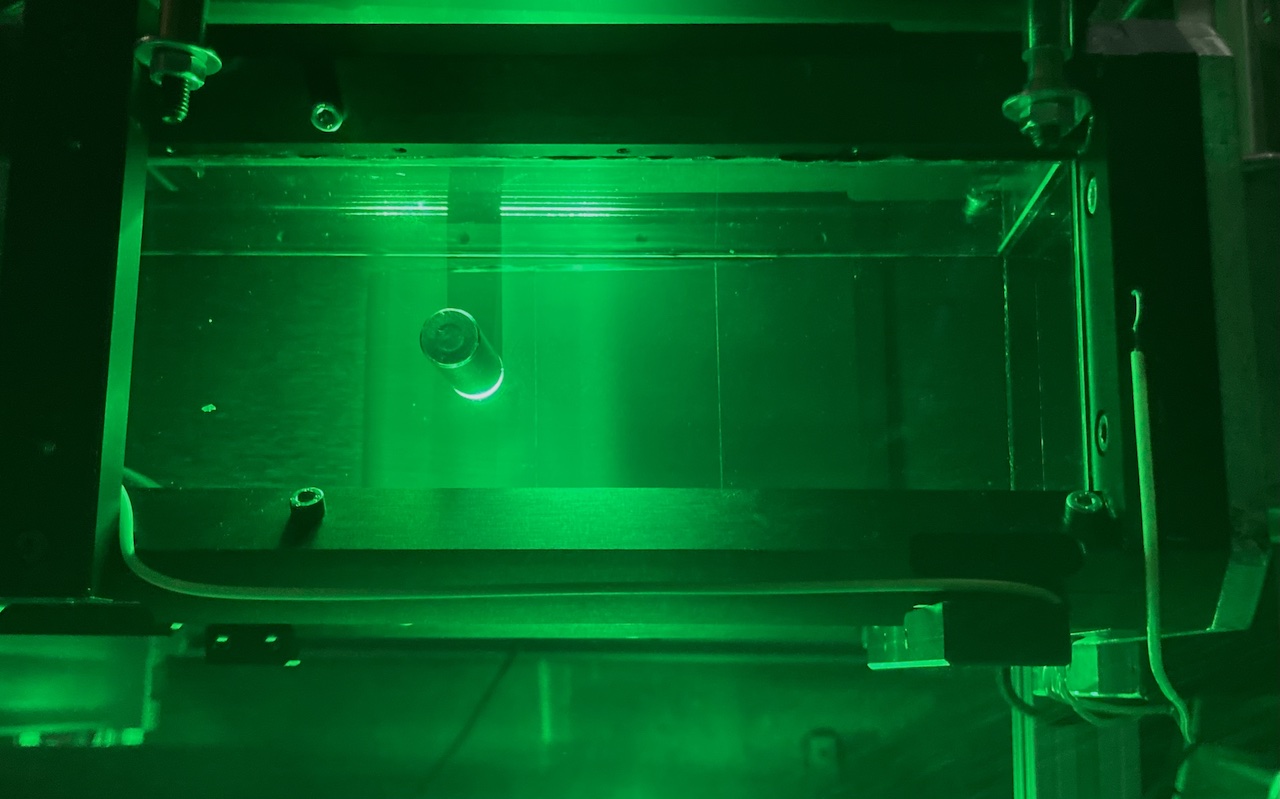}\\[2mm]
    \includegraphics[width=0.94\columnwidth]{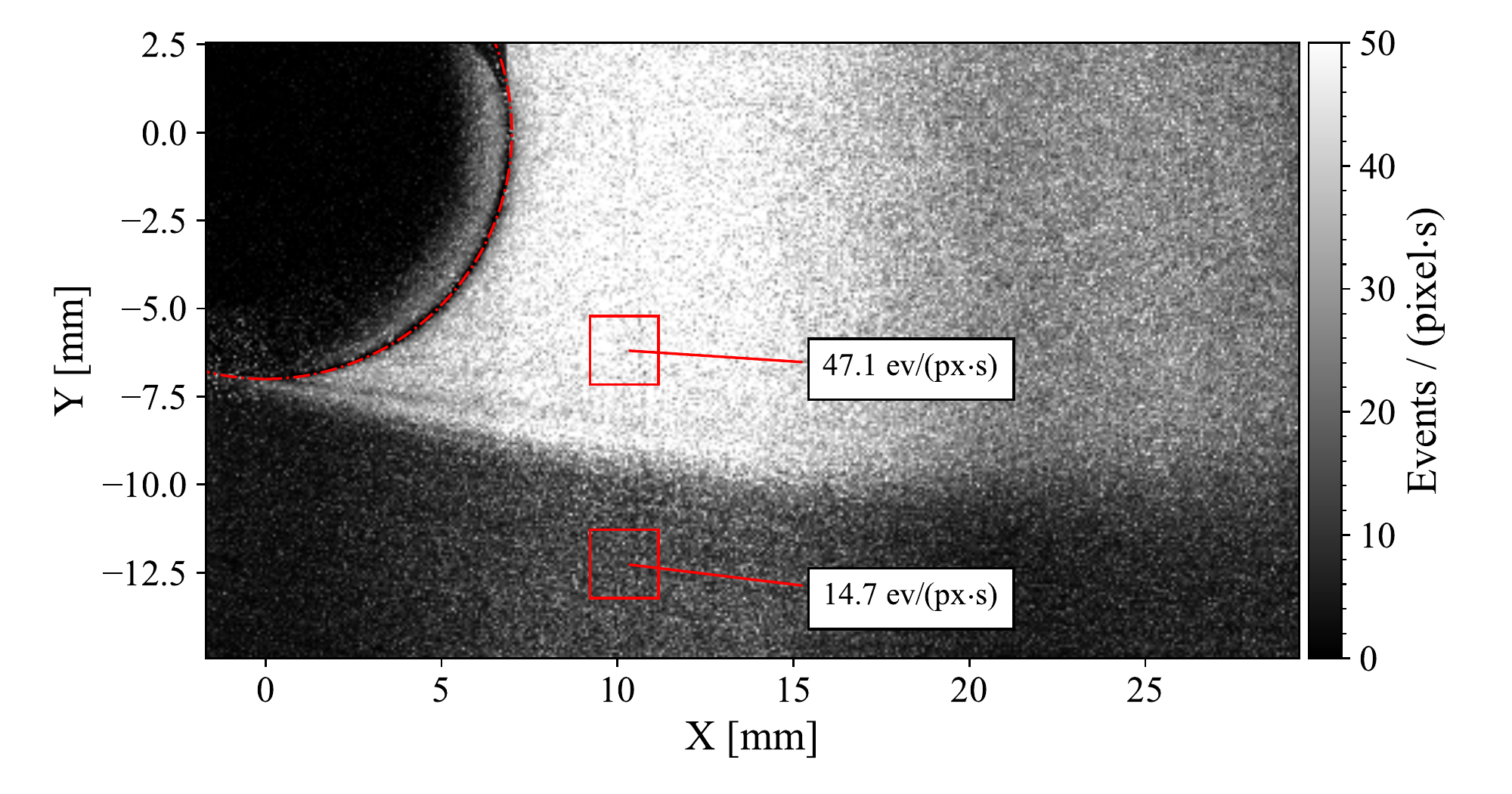}\\
    \includegraphics[width=0.69\columnwidth]{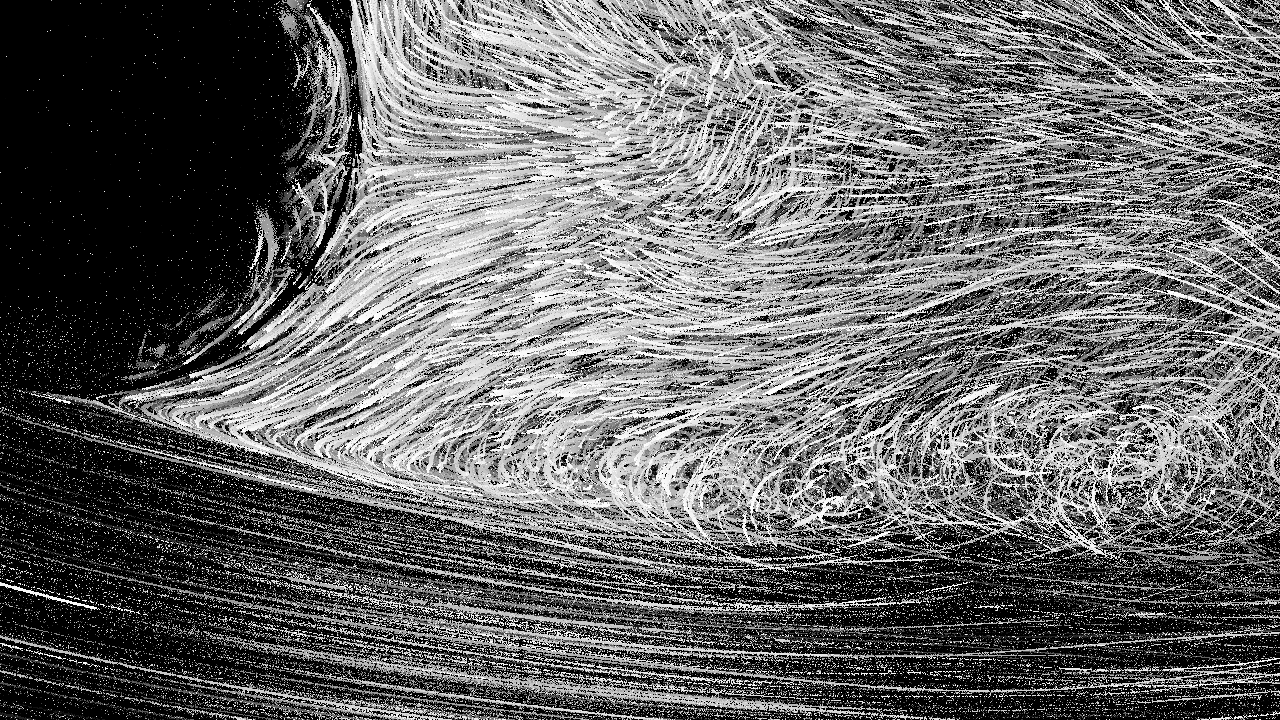}
    \caption{Top: Photograph of imaging setup for flow around a cylinder of 14\,mm diameter within a square duct of 76\,mm side length. Droplet based aerosol seeding of 1\,$\mu$m is present; the shadow of the cylinder within the light sheet is visible.
    Middle: Event recording rate within the imaged domain. The cylinder surface is outlined with the red-dashed circle.
    Bottom: Recorded particle event tracks for a duration of 50\,ms highlighting the surface of the cylinder.
    The outer flow has a velocity of 1.7 m/s; magnification is 0.0237\,mm/pixel. % (42.1\,pixel/mm).
    }
    \label{fig:airflow7_exp}
\end{figure}

Figure~\ref{fig:airflow7_exp} shows a photograph of measurements performed in a small wind tunnel with a with a square cross section of 76$\times$76 mm$^2$ and 1.5\,m length. A steel cylinder of 14\,mm diameter spanning the height of the channel is placed on the centerline.
At a bulk flow velocity of $U_b \approx 1.5$\,m/s the diameter-based Reynolds number is $\mathrm{Re}_D = 1400$, which is well beyond the transition to turbulence resulting in a turbulent wake.
At the upstream end of the wind tunnel the flow is seeded with 1$\mu$m paraffin droplets produced by an atomizing aerosol seeder (Laskin type). The droplets are illuminated by a CW laser with 5.6\,W maximum power (Kvant Laser, $\lambda$ = 520\,nm) using a light sheet with a waist thickness of about 1\,mm.

Measurements are performed at different magnifications and fields of view.
Here, we will focus on close-up measurements of the shear layer formed downstream of the flow separation on the cylinder surface. This area is characterized by high dynamics within the flow and includes quiescent regions on the downstream side of the cylinder and strong velocity gradients within the shear region. Flows such as these are challenging to capture using planar flow measurement techniques.
At a magnification $m = 41.2$\,pixel/mm the imaged area covers 31.0$\times$17.5\,mm$^2$.
The light sheet is aligned with the centerline of the test section and passes from bottom to top in the photograph of Figure~\ref{fig:airflow7_exp}, while the flow moves from left to right.

Figure~\ref{fig:airflow7_exp}, middle, shows an image of the cumulative intensity change events and represents a total of 1.3$\cdot$10$^8$ collected in 6 seconds.
The event rate, measured in events per pixel and time, varies by about a factor of 3 between the quiescent wake and the faster outer flow in spite of the fact that the flow throughout the field of view in homogenously seeded (Also the wake does not appear brighter in the photograph).
An explanation for this effect may be the limited latency (typ. $\approx$100~$\mu$s, see Table~1 in \cite{EBVreview:2022}) of the event camera sensor: particles in the outer flow either move too fast or are too faint to trigger the contrast detection circuit on the pixel level that result in intensity-change events.
With increased speed of the intensity gradient moving across the pixel (e.g. illuminated particle), the pixel will have a reduced likelihood of responding to the change. The event generation is a probabilistic process that has a stochastic nature depending a number of parameters \cite{EBVreview:2022,HuDelbruck:2021}. Among the faster moving particles in the outer flow it is only the brighter (larger) ones that trigger events.

Noteworthy is the absence of laser flare in the event recordings such as on the cylinder itself.
In the particle streak image of Figure~\ref{fig:airflow7_exp}, bottom, the outline of the cylinder can be identified by the mirrored particle image event tracks on ``inside" the cylinder. Pixels on the cylinder surface produce no events which is either due to time-constant intensity levels or due to saturation of the photodiodes. Image-blooming, which is common for overexposed CCD sensors, cannot be observed.
The fact that no stationary ``particles" are present on the surface allows the motion of actual particles to tracked within a few pixel distance of the surface which in itself has been challenging task for the PIV technique.

Compared to the water flow, the air flow has a much higher ``pixel velocity" exceeding 50\,000\,pixel/s. This is about four times higher than the measurements performed with the rotating plate of particles (c.f. Figure~\ref{fig:solidbody_data}).
In their work on adaptive time-slice block-matching optical flow algorithms (ABMOF) Liu \& Delbruck \cite{Liu:2018} quote values of $>$30\,000 pixel/s as ``extremely fast".
The presently available processing algorithms based on motion compensation and sum-of-correlation have been found not suitable to reliably recover velocity information simultaneously in the slow wake region and the fast outer flow.
In order to capture the outer flow, the single pass motion compensation algorithm requires a sampling area exceeding 100 pixel. On the other hand the wake region can be reliably processed with samples of 40$\times$40\,pixel (1.0$\times$1.0\,mm$^2$).
Similarly, the temporal offset $\tau$ of the sum-of-correlation approach needs by sufficiently short in order to capture the event tracks of fast moving particles in both sampling volumes.
To properly capture the high dynamic range of velocities, an adequate processing strategy first needs to be developed.

\subsection{PIV processing of the air flow event data}
\label{sec:airflow-proc}
Through temporal sampling of the event-data, image sequences can be generated that can then be processed with conventional, cross-correlation based PIV algorithms. The present data of the near-cylinder shear layer is sampled for a duration of $T$ = 250\,$\mu$s at intervals of $\Delta T$ = 250\,$\mu$s. With this sampling, the outer flow exhibits a particle image displacement magnitude of about 20 pixels for a velocity magnitude of 1.7\,m/s.
The resulting image sequence is processed with the previously mentioned multi-frame, grid-refining algorithm using final sampling windows of 64$\times$64 pixel to accommodate the rather low event density and correspondingly low particle image density of the outer flow.

\begin{figure}[htb]
\centering
    \includegraphics[width=0.49\columnwidth]{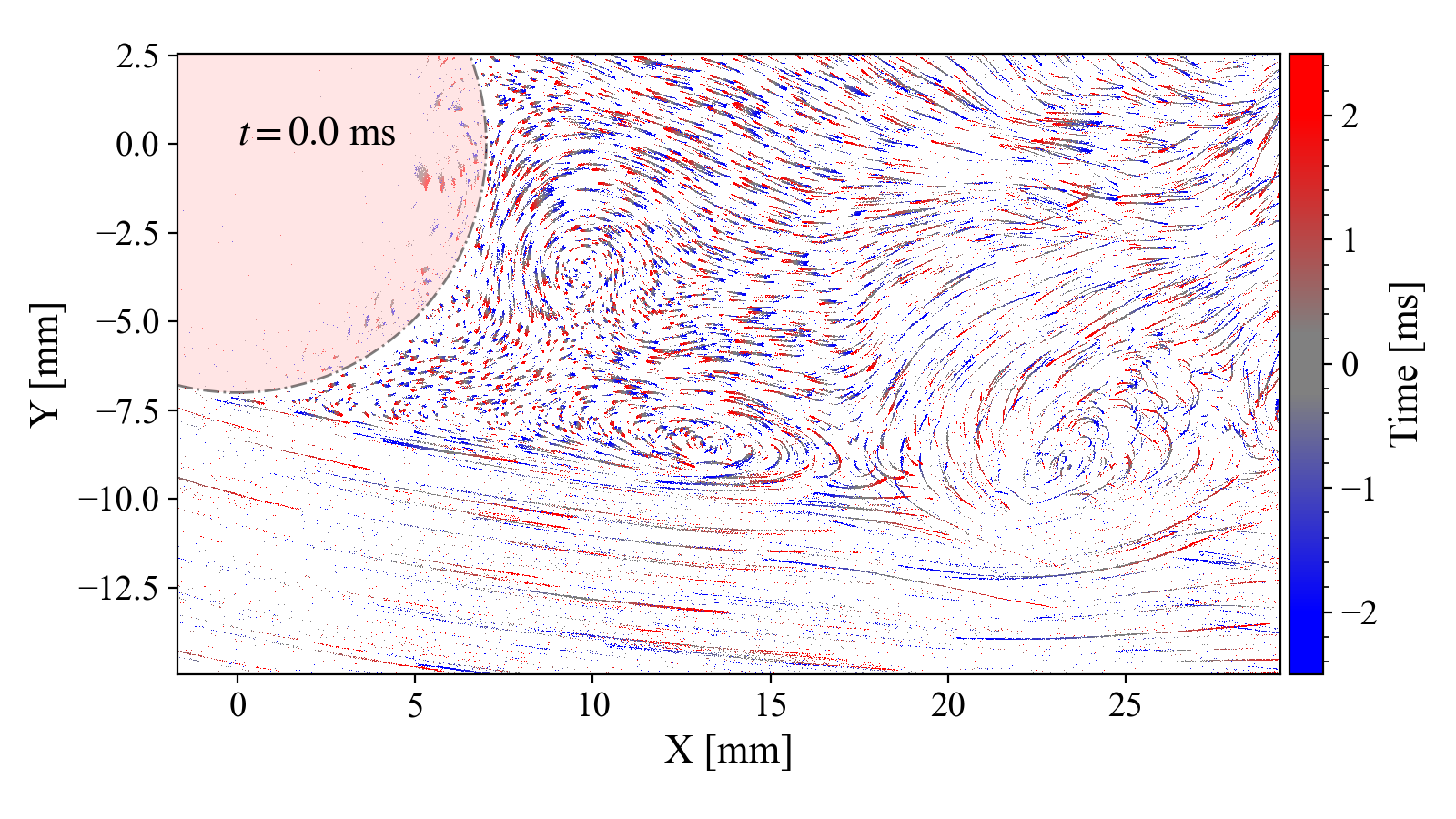}
    \includegraphics[width=0.49\columnwidth]{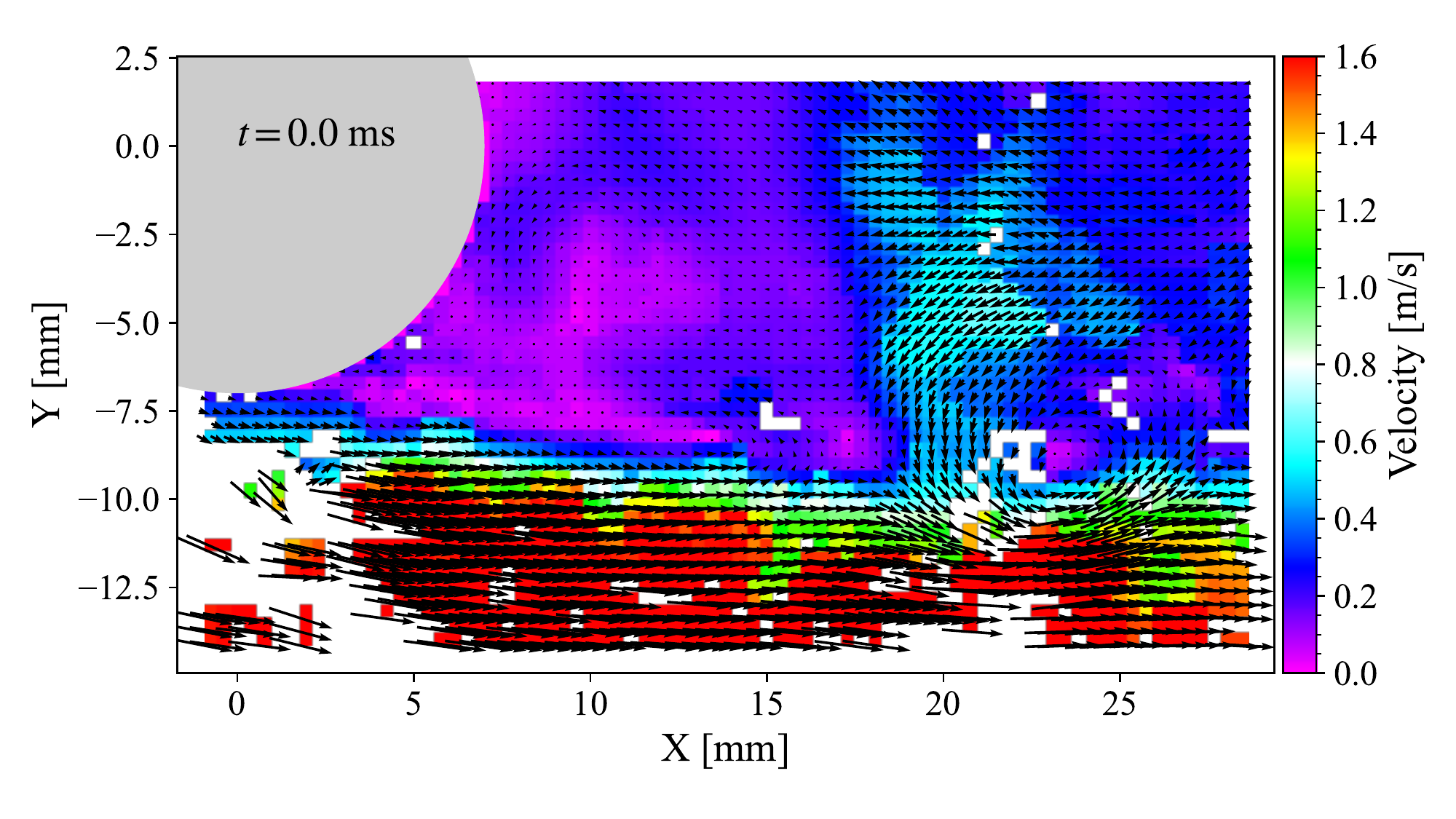}\\
    \includegraphics[width=0.49\columnwidth]{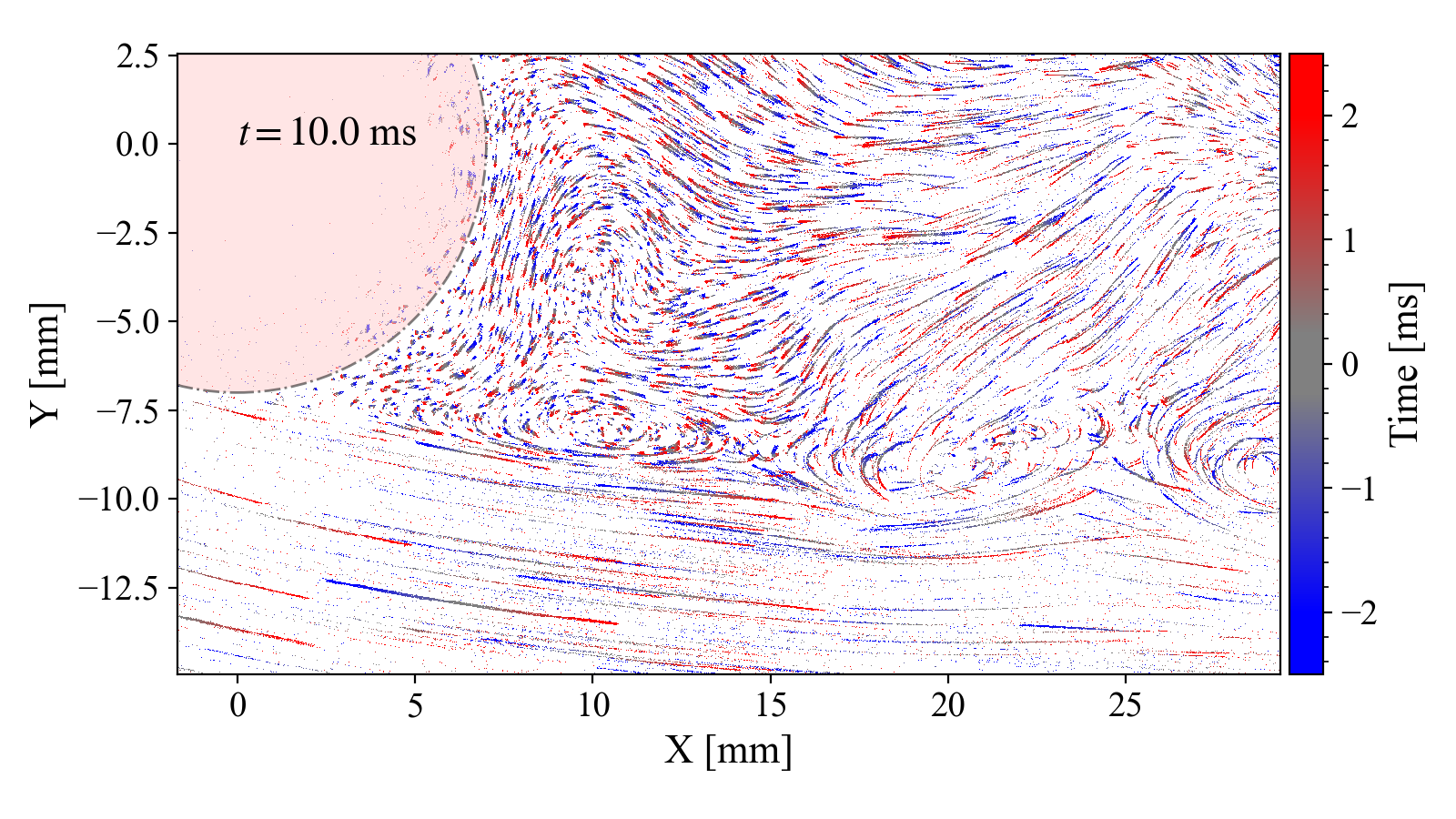}
    \includegraphics[width=0.49\columnwidth]{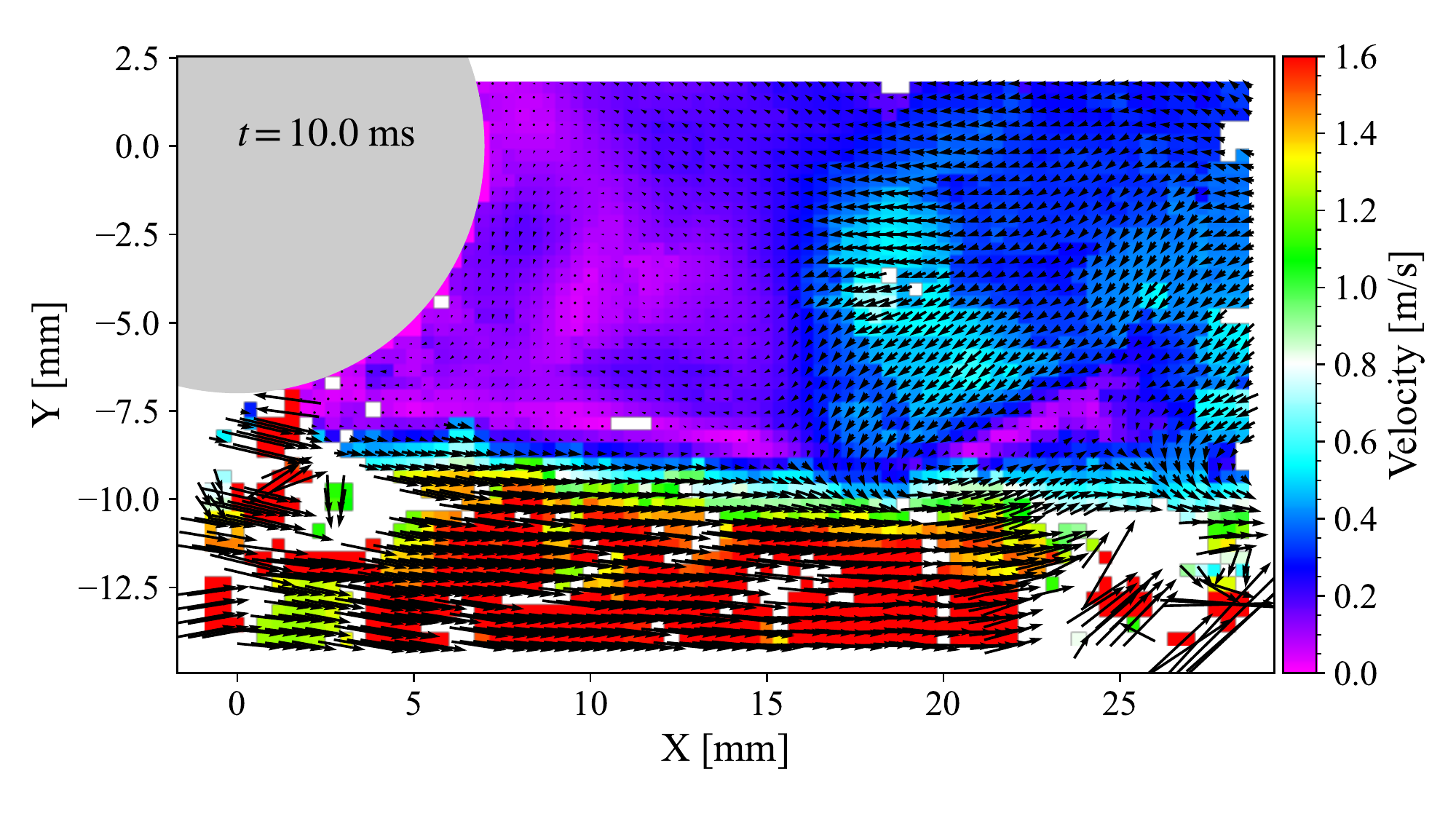}\\
    \includegraphics[width=0.49\columnwidth]{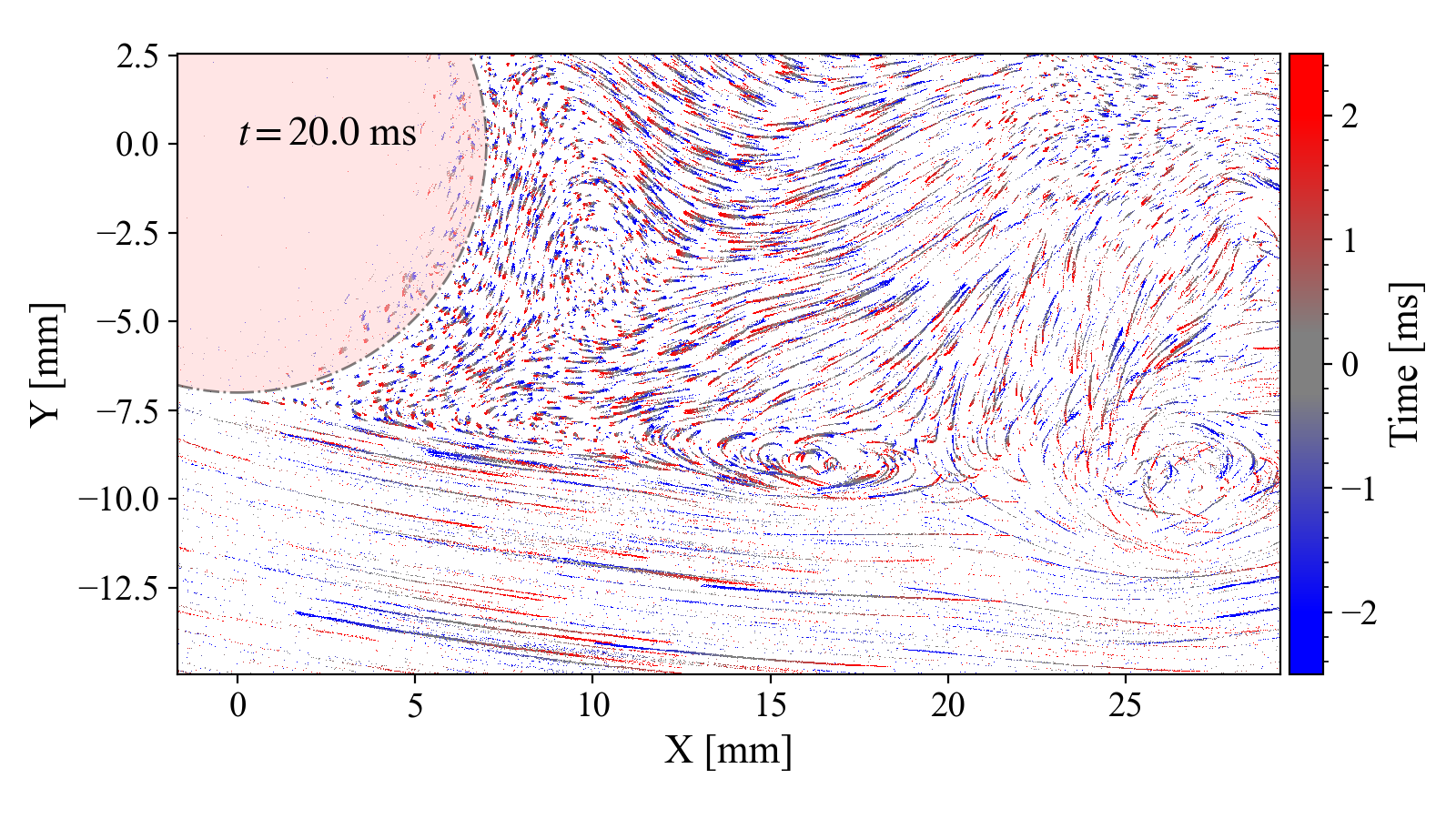}
    \includegraphics[width=0.49\columnwidth]{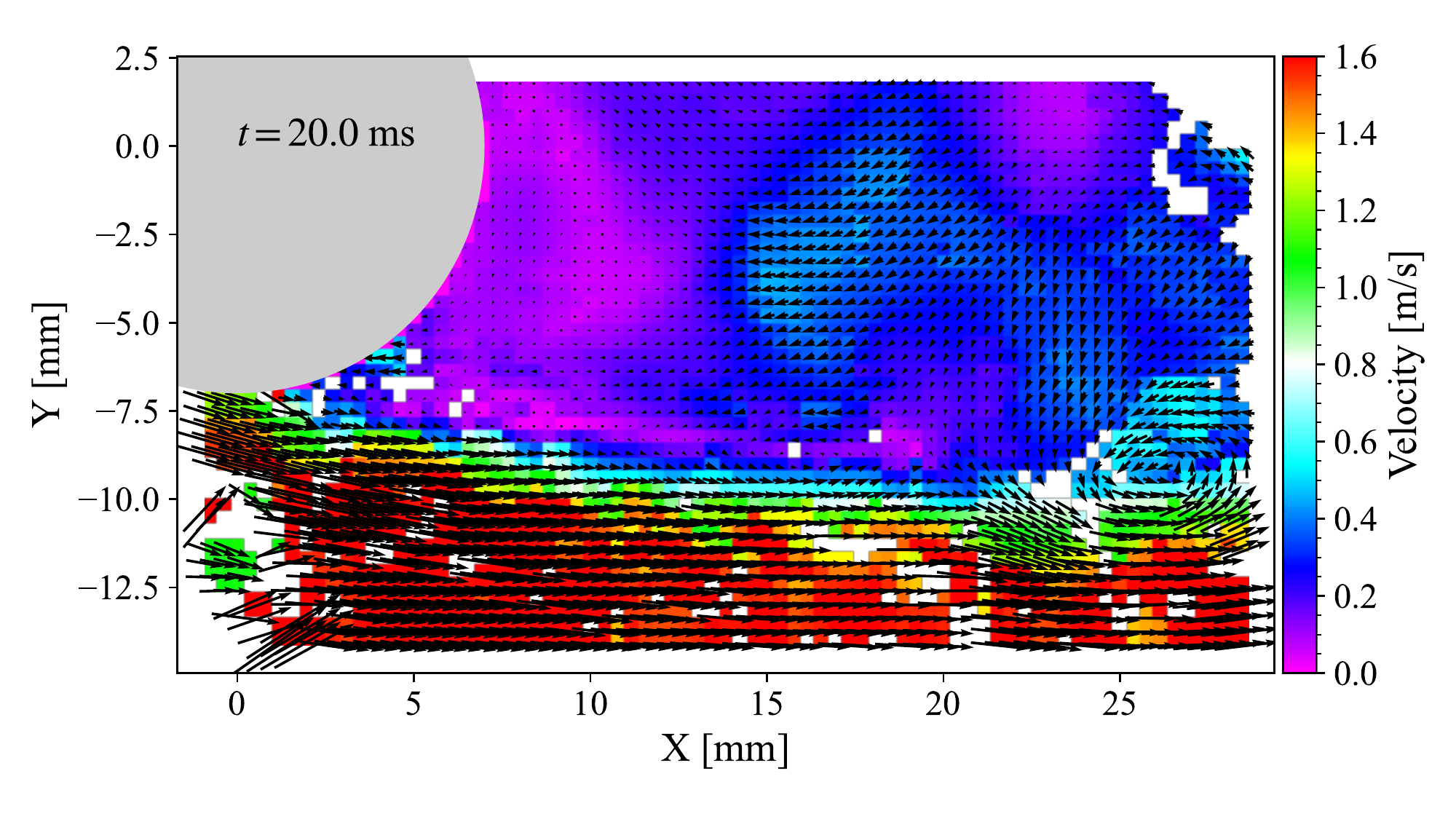}\\
    \includegraphics[width=0.49\columnwidth]{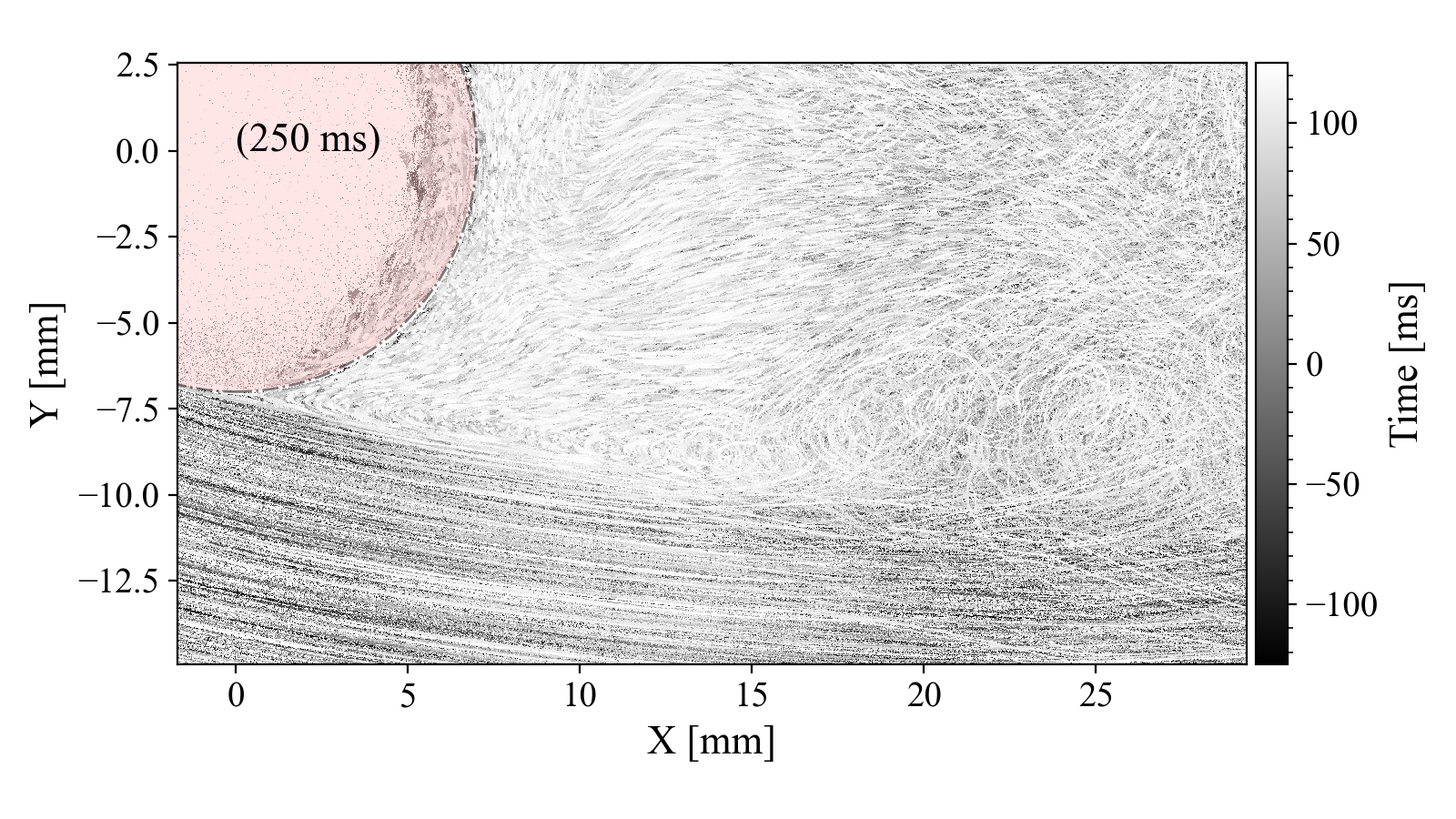}
    \includegraphics[width=0.49\columnwidth]{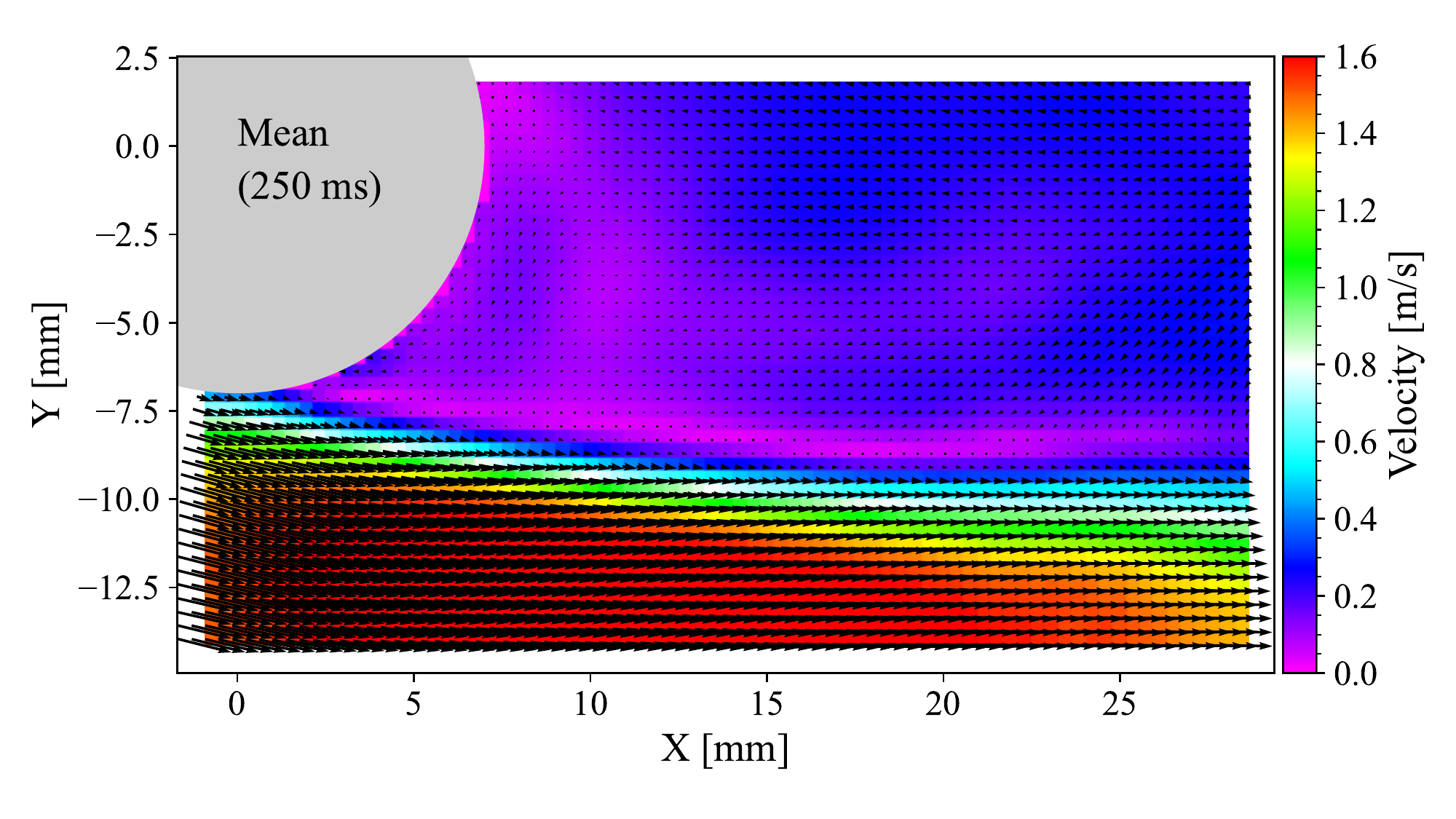}
    \caption{%
    Left column: Event-images of the near cylinder shear layer for different times.
    Right column: velocity maps obtained with PIV processing of the event data sampled for time intervals of $T$ = 250\,$\mu$s.
    Bottom row: shows both the event data and mean flow field for a duration of 0.25\,s.
    }
    \label{fig:airflow7_data}
\end{figure}

Selected velocity maps from a sequence of 1000 ``reconstructed" PIV recordings are provided in Figure~\ref{fig:airflow7_data} alongside with corresponding particle event tracks.
The high flow velocity and reduced event density of the outer flow results in considerable data dropout, whereas very high validation rates are achieved within the cylinder wake. The recovered data set has a temporal resolution of 4\,kHz albeit lowpass filtered by the multi-frame PIV processing scheme.
An adaptive processing scheme is required to further bring out the fine-scale structure of the immediate cylinder wake without compromising the velocity estimates from the outer flow obtained at coarser resolutions.

\section{Discussion}
\label{sec:discussion}
From the previous experiments image velocities on the order of $\pm$10\,000\,pixel/s were found feasible with the present event camera hardware.
Given a field of view 100\,mm, that is imaged by a sensor of 1000 pixels side length, flow velocities in range of ($\pm 10$\,pixel/ms) $\times$ (100\,mm / 1000\,pixel) = $\pm$1\,m/s are possible.
By increasing the field of view to 1\,m, the measurements range is linearly increased to $\pm$10\,m/s.
As these limitations seem to be rooted in the sensors' latency, it will require advancements on the detector end to extent to velocity range to higher values.

While it is true that an event camera actually produces a reduced number of data while viewing a scene, in particular one with limited dynamic content, the imaging of many small and constantly moving particles quickly leads to a considerable amount of incoming events.
The investigated water and air flows produced on the order of 20$\times$10$^6$ events/s resulting in data rates of up to 60~MB/s. With increased seeding densities the data rates increase proportionally, resulting in data drop-out in the limited bandwidth data acquisition channel.
During the previously described experiments the seeding density and directly coupled event data rate had to be carefully adjusted in order not to cause data-dropout, which was accompanied with loss of time-stamp information and complete loss of signal.

Some of the particle event tracks of the recorded air flow in Figure~\ref{fig:airflow7_exp} have a width of only 1 pixel which for PIV would result in so-called \emph{pixel-locking} artifacts. Possible influences of this and other effects on the measurement uncertainty of EBIV will have to be investigated further.

The following summarizes both the advantages and disadvantages of flow velocimetry using event-based cameras.

\noindent\textbf{Advantages:}
\begin{itemize}
    \item Event-based sensing is suitable for real-time flow visualization. % the recorded data of particle movement is visually simple and intuitively to comprehend.
The event tracks produced by the movement of the particle field directly visualize the flow field without the need of further processing.
    \item Event-based cameras are tolerant against intensity variations within the light sheet because its sensors only react on intensity changes.
    \item Event-based cameras are tolerant against laser flare (light scatter on surfaces) and non-uniform background intensity.
    \item Event-based cameras are tolerant against high background intensity. This allows particles to be tracked in bright environments, assuming this intensity is essentially constant in time (i.e. varies slowly in time).
    \item Particle event tracks are clearly distinguishable from the background and can be extracted either by feature detection or by using the clusters formed in the motion-compensation analysis.
    \item Both in-plane and out-of-plane loss of particle images is not as critical as for PIV. For particles leaving the sample volume the number of events contributing to the velocity estimate is reduced which results in an increased uncertainty.
    \item The time-resolved nature of the event data can be used to extract additional information of the particle motion such as acceleration or path-curvature. Contrary to pulse-illuminated PIV (or PTV) image frames, the event-trails generated by moving particles provide continuous information on their location with microsecond resolution.
    %\item Processing of event-based image data is less complex as for PIV and can be highly parallelized.
    \item Conventional PIV processing is possible by generating sub-frames from the time-resolved event data.
    \item Hardware requirements are less demanding in comparison to conventional PIV systems relying pulsed light illumination (see below).
\end{itemize}

\noindent\textbf{Disadvantages:}
\begin{itemize}
    \item Slow moving particles will produce less events per given time. In areas of slow or nearly quiescent flow, an increase of the sampling time will result in more events from which to reconstruct the velocity. On the other hand, faster moving particles have a reduced likelihood of producing events which is due to the inherent latency of the EBV sensor (typ. O(100\,$\mu$s)).
  \item Temporal intensity variations (such as 50\,Hz lamp flicker) within the image scene will trigger many simultaneous events that are not related to the motion of objects. This can be addressed through the design of sensor using e.g. the \emph{Global Hold, Global Reset} approach (see \cite{Ryu:Samsung:2019}).
      The Gen.4 sensor of evaluation kit by Prophesee has an anti-flicker filter option available.
      (This spontaneous generation of events could be harnessed to briefly make stationary or slowly moving particles visible thereby overcoming the previously mentioned shortcoming.)
  \item The velocity range is restricted which limits the applicability of EBIV to slow flows. While for PIV the velocity range can be adapted by simply changing the separation of the laser pulses (or frame rate of the high-speed camera), no such option exists for EBIV. The upper bound on reliable velocity detection is on the order of 50\,000 pixel/s and seems due to sensor's latency with present day event camera hardware.
  \item The bandwidth of EBV hardware can be a limiting factor as the number of events scales linearly with seeding density (i.e. illuminated particles).
\end{itemize}

\section{Conclusion \& Outlook}
\label{sec:concl}
The material presented in the preceding sections constitutes a feasibility study on the use of the EBV for the estimation of fluid flow velocity maps. The work is focussed on obtaining dense 2d-2c data using particle image densities previously not reported with event-based imaging. While there is no intention in rivaling the well established PIV techniques, which follow frame based approaches, event-based imaging velocimetry (EBIV) may nonetheless provide some attractive advantages.
The flow measurements clearly demonstrate that EBIV enables measurements in close proximity of surfaces where PIV and other frame-base imaging methods generally have problems with excessive light scatter.
As surfaces generally do not move, they do not trigger any intensity-change events and hence are not visible by the event camera (this also makes calibration using stationary targets a little more challenging).
Another advantage is that individual particles can be identified efficiently using the contrast maximization approach used for the recovery of the motion field.
Although not implemented in the present work, the continuous event tracks produced by the imaged particles allows the continuous reconstruction of the particle paths in time and 2d space.
As demonstrated in previous works \cite{Borer:2017,Wang:2020}, this can be readily extended to time-resolved 3d particle tracking.

From an implementation point of view, the investment cost is reasonable, given the fact, that instead of a pulsed laser, a low-cost CW laser can be used. The sensitivity of event cameras is considerably higher such that moderate laser powers in the 1-5 Watt range are already sufficient to adequately image micrometer-sized particles in water and air flows with viewing domains of 10\,cm side length.
%The high sensitivity also make the use of high-power LEDs for illumination becomes viable \cite{Willert:PulsedLED:2010}.
Beyond this, EBIV does not require elaborate synchronization between camera and laser.
The currently available event camera hardware can be synchronized between different units for multi-view imaging and can additionally record external synchronization events in the data stream.
Finally, unlike specialized cameras used for PIV (e.g. actively cooled sensor, double-shutter capability), event cameras are available at a significantly reduced unit cost as their use is intended for a much wider range of applications such as autonomous navigation, 3d sensing, object counting, to name a few.

Processing of the captured event data can be optimized considerably through GPU implementations such that real-time processing becomes feasible. Latency between data acquisition (event data input) and result can be on the order of milliseconds, as already demonstrated in a multitude of applications using EBV (e.g. real-time autonomous  navigation \cite{Rebecq:2018,Kim:2021,Liu:2021} or collision avoidance \cite{Falanga:2020}).

EBV sensor development is ongoing with some event cameras already employing sensors with megapixel (HD) resolution along with small pixels in the 5$\,\mu$m range. The use of back-side illuminated pixels, such as in the device used in the current study, further increases sensitivity and dynamic range. Currently the systems can image at an equivalent frame-rate of up to 5-10~kHz.
A reduced latency of the EBV sensors could extend the use of event-based vision to capture even faster events.
The previously shown velocity data of the cylinder wake flow has an effective temporal resolution of 250\,$\mu$s with a corresponding frame rate of 4~kHz.

Beyond the application described herein, event-based vision will certainly find other applications in the field of flow diagnostics and related areas such as time resolved background oriented schlieren (BOS) and flow visualization.

\section*{Supplementary information}
\small{Animations of the raw event data and reconstructed velocity field shown in Figures~\ref{fig:events_image} and \ref{fig:eval_onepass} are provided as supplementary files.}

\section*{Declarations}
\small{
\begin{description}
  \item[Author Contributions] C.W. wrote the manuscript and code, J.K. reviewed the work and contributed via discussions. C.W and J.K. both conducted the experiments.
  \item[Funding] This research received no external funding.
  \item[Institutional Review Board Statement] Not applicable.
  \item[Informed Consent Statement] Not applicable.
  \item[Data Availability Statement] Given the exploratory nature of the presented work the authors conclude that the work contains no data of archival relevance. The raw data of the water and air flow can be provided upon request (a dedicated Github site is under consideration).
  \item[Code Availability Statement] Python code for basic EBIV analysis can be provided by the author upon request (a dedicated Github site is under consideration).
  \item[Conflict of Interest] The authors declare no conflict of interest.
\end{description}}

\bibliographystyle{unsrtnat}
\bibliography{EBIV_lit} 

\end{document}